\documentclass[11pt,a4paper,final]{article}
\pdfoutput=1
\usepackage[margin=1in, top=1.2in, bottom=1.2in]{geometry}

\usepackage[T1]{fontenc}

\usepackage{authblk}
\usepackage{amsmath}
\usepackage{amssymb}
\usepackage{braket}
\usepackage{bbold}
\usepackage{dsfont}

\usepackage{dcolumn}
\usepackage{makecell}
\usepackage{multirow}

\usepackage{float}
\usepackage{caption}
\usepackage{subcaption}

\usepackage{verbatim}
\usepackage{tikz}
\usetikzlibrary{positioning,shapes}

\begin{document}
\title{D=4 supersymmetric Yang-Mills quantum mechanics with three colors}
\author{Zbigniew Ambrozi\'nski
\thanks{zbigniew.ambrozinski@uj.edu.pl}
}
\affil{Jagiellonian University, Lojasiewicza 11, Krakow, Poland}

\maketitle
\begin{abstract}{The spectrum of the $D=4$ supersymmetric Yang-Mills quantum mechanics with $SU(3)$ gauge group symmetry is computed in different channels with definite total angular momentum and the total number of fermions. In sectors with small number of fermions the spectrum is shown to be purely discrete. Continuous spectrum appears in channels with more fermions. Supersymmetry generators are used to identify supermultiplets and determine the level of restoration of supersymmetry for a finite cutoff.}
\end{abstract}


\section{Introduction}

In this paper we present a detailed study of the supersymmetric Yang-Mills quantum mechanics (SYMQM) with $SU(3)$ gauge group \cite{Claudson,Ferrara,Brink}. The addressed system is dimensionally reduced $D=4$ supersymmetric Yang-Mills gauge theory. The reduced model has four supercharges. It belongs to a broader class of SYMQM with various dimensions $D=2,4,6,10$ and different symmetry groups $SU(N)$. In the past years many such systems were studied. Bosonic part of SYMQM in $D=4$ with $SU(2)$ and $SU(3)$ gauge symmetry was considered as a zeroth order approximation to the small volume expansion of the theory of gauge fields in QCD \cite{Bjorken,Munster,Weisz,Ziemann_phd,Koller1,Koller2}. Later, the $SU(N)$ theory was given an interpretation of a non-perturbative description of the supermembrane \cite{Hoppe,Bergshoeff,deWit}. There, $N$ plays a role of a cutoff and the theory in continuum is reproduced in the large $N$ limit. However, it was shown that such supermembranes are unstable \cite{Nicolai,Helling} which was considered as a large setback. Later the BFSS conjecture, which relates SYMQM in the large $N$ limit and $D=10$ to the dynamics of D0 branes in M-theory, was proposed \cite{BFSS}. The BFSS conjecture aroused a large interest in this model \cite{Becker,Porrati,Taylor}. Asymptotic ground states for the $SU(2)$ case were found in \cite{Danielsson,Halpern}. The lattice methods were applied in \cite{Catterall,Wiseman}. SYMQM was also studied with hybrid Monte Carlo approach \cite{Anagnostopoulos,Hanada,Nishimura}.

A program to study the whole family of SYMQM with various dimensions and gauge groups was proposed in \cite{Janik,Wosiek}. The cut Fock space method, which will be our tool was used in  a set of papers. In  \cite{Wosiek,Campostrini} the $D=4$ model with $SU(2)$ symmetry group was addressed. The $D=2$ case was studied and eventually a complete solution for all $SU(N)$ was found in \cite{Trzetrzelewski_susyd2,Korcyl,KorcylN}. There was yet another study based on the Schr\"odinger equation \cite{vanBaal,Kotanski,Kotanski2}. With this method the energies and eigenstates of $D=4$ model with $SU(2)$ gauge group were found with great precision in the sector with 0-angular momentum and two fermions.

In this paper the system is solved with the cut Fock space method. The idea of this approach is to represent the Hamiltonian as an infinite matrix in the Fock basis and to introduce a cutoff. The cutoff limits the total occupation number of bosons. Thanks to conservation of number of fermions, the Hamiltonian can be considered in each fermionic sector separately. Construction of the matrix is performed using a recursive algorithm. This is the most numerically involved part of calculations. Once the matrix is created, it is diagonalized to obtain its eigenvectors and eigenvalues.

In the central part of this paper we analyze the spectrum in channels with definite number of fermions and angular momentum. An important question is to distinguish the discrete spectrum from the continuous one. This analysis is based on dependence of energies on the cutoff. It turns out that the spectrum in the few lowest fermionic channels is discrete, while the continuum spectrum appears for many fermions. This conclusion was already found for the case of $SU(2)$ \cite{Wosiek,Campostrini} and is now confirmed also for $SU(3)$.

Finally, the supersymmetry multiplets are identified and discussed. For a finite cutoff the supersymmetry is broken and there are no degenerate supermultiplets. In order to identify states which form multiplets in the continuum limit, we define so called supersymmetry fractions. They measure how supercharges map some energy eigenstates into other. Values of supersymmetry fractions in the continuum limit are known. Therefore, they are useful to analyze the degree of breaking of supersymmetry. Moreover, they help to identify SUSY partners.

\section{The model}\label{ch:model}

The supersymmetric Yang-Mills quantum mechanics is obtained by dimensional reduction of the Yang-Mills field theory \cite{Claudson}. The full space is then reduced to a single point. Therefore, the gauge field $A_\mu(t,\vec x)$ depends only on $t$. We work in the temporal gauge $A_0=0$ and adopt the notation of quantum mechanics $A_i=x_i$. Finally, SYMQM with $SU(3)$ gauge group in four dimensions is described by the Hamiltonian
\begin{align}\label{eq:hamiltonian}\begin{split}
H&=H_K+g^2H_V+gH_F=\frac{1}{2}p^a_ip^a_i+g^2\frac{1}{4}f^{abc}f^{ade}x^b_ix^c_jx^d_ix^e_j+g\frac{i}{2}f^{abc}\psi^{a\dagger}_\alpha(\Gamma_k)_{\alpha\beta}\psi^b_\beta x^c_k.
\end{split}\end{align}
Spatial indices $i,j,k$ take values $1,2,3$ and color indices $a,b,c,d,e$ range from $1$ to $8$.
$\Gamma_k=\gamma^0\gamma^k$ are the alpha matrices.
Bosonic position and momentum operators $x_i^a$ and $p_i^a$ satisfy canonical commutation relations $[x_i^a,p_j^b]=i\delta^{ab}\delta_{ij}$ whereas $\psi$ is a Majorana spinor satisfying $\{\psi_\alpha^a,\psi_\beta^{b\dagger}\}=\delta^{ab}\delta_{\alpha\beta}$.

Physical states are gauge singlets. That is, the Hilbert space is limited to states $\ket{s}$ for which $G^a\ket{s}=0$, where
\begin{align}
G^a&\equiv f^{abc}(x^b_ip^c_i-\frac{i}{2}\psi^{b^\dagger}_\alpha\psi^c_\alpha)
\end{align}
are generators of $SU(3)$.

In what follows in all explicit calculations the Weyl representation of Dirac matrices from \cite{Itzykson} will be used. In this representation
\begin{align}
\gamma^0&=\left(\begin{array}{cc}0&\mathds 1\\\mathds 1&0\end{array}\right),&
\gamma^k&=\left(\begin{array}{cc}0&\sigma_k\\-\sigma_k&0\end{array}\right),&
\gamma^5&=\left(\begin{array}{cc}-\mathds 1&0\\0&\mathds1\end{array}\right).
\end{align}
The Majorana condition for $\psi$ reads $\psi^a_C=\psi^a$ where
\begin{align}\label{eq:Majorana_conjugate}
\psi^{a}_C&=C(\bar{\psi}^a)^T,&C&=\left(\begin{array}{cc}-i\sigma_2&0\\0&i\sigma_2\end{array}\right).
\end{align}
It is convenient to express $\psi$ in terms of a two component complex spinor $f_\alpha$ as
$\psi^a=\left(f^{a}_1,f^{a}_2,f^{a\dagger}_2,-f^{a\dagger}_1\right)^T$  \cite{Weinberg},
where fermionic operators are defined to satisfy canonical anticommutation relations
\begin{align}\begin{split}
\{f_\alpha^{a\dagger},f_\beta^{b}\}&=\delta^{ab}\delta_{\alpha\beta},\\
\{f_\alpha^{a},f_\beta^{b}\}=\{f_\alpha^{a\dagger},f_\beta^{b\dagger}\}&=0.
\end{split}\end{align}
The Majorana condition and anticommutation relations for $\psi$ follow automatically.

In what follows the matrix notation will be used. For an operator $\mathcal O^a$ with an adjoint $SU(3)$ index we define $\mathcal O=\mathcal O^aT^a$ where $T^a$ are generators of $SU(3)$ in the fundamental representation. We use normalization in which multiplication law of $T^a$ reads
\begin{align}
T^aT^b=\frac{1}{6}\delta^{ab}\mathbb 1_3+\frac{1}{2}(if^{abc}+d^{abc})T^c.
\end{align}
In the matrix notation the Hamiltonian has the form
\begin{align}
H&=Tr\left(p_ip_i-\frac{1}{2}[x_i,x_j]^2+(\Gamma_k)_{\alpha\beta}\psi^\dagger_\alpha[\psi_\beta,x^k]\right).
\end{align}

\subsection{Symmetries}
We come to discussion of symmetries of the Hamiltonian. The supersymmetry gives rise to supercharges which are given by
\begin{align}\label{eq:supercharges}
Q_\alpha=(\Gamma_k\psi^a)_\alpha p^a_k+igf^{abc}(\Sigma_{jk}\psi^a)_\alpha x^b_jx^c_k,
\end{align}
where $\Sigma_{jk}=-\frac{i}{4}[\Gamma_j,\Gamma_k]$.
They obey anticommutation relations
\begin{align}\label{eq:anticommutation_relations}
\begin{split}
\{Q_\alpha,Q^\dagger_\beta\}&=2\delta_{\alpha\beta}H+g(\Gamma_k)_{\alpha\beta}x^a_kG^a.
\end{split}
\end{align}
In the space of gauge singlets the anticommutator $\{Q_\alpha,Q^\dagger_\beta\}$ is proportional to $H$.

The system has rotational symmetry inherited from the field theory. That is, the Hamiltonian commutes with angular momentum operators which are given by
\begin{align}\label{eq:angular_momentum}
J_i=L_i+S_i=\epsilon_{ijk}\left(x^a_jp^a_k+\frac{1}{4}\psi^{a\dagger}\Sigma_{jk}\psi^a\right).
\end{align}

A remarkable feature of the four-dimensional theory is conservation of the total number of fermions $n_F=f^{a\dagger}_\alpha f^{a\dagger}_\alpha$.  This symmetry allows one to study each sector with given $n_F$ independently, which is useful in practice. Because of the Pauli exclusion principle there are $17$ sectors with $n_F=0,\ldots,16$. This symmetry is a property only of $2$ and $4$-dimensional theories. It is not present e.g. in 10 dimensions which is interesting due to the BFSS conjecture.

Furthermore, there is a particle-hole symmetry:
\begin{align}\label{eq:ph_symmetry}
f^a_1&\rightarrow -f^{a\dagger}_2,&f^{a\dagger}_1&\rightarrow -f^{a}_2,\nonumber\\
f^a_2&\rightarrow f^{a\dagger}_1,&f^{a\dagger}_2&\rightarrow f^{a}_1,\\
x_i^a&\rightarrow -x_i^a,&p_i^a&\rightarrow -p_i^a.\nonumber
\end{align}
A natural consequence of this symmetry is that one can find eigenstates of the Hamiltonian which are even or odd under (\ref{eq:ph_symmetry}). On the other hand, conservation of $n_F$ implies that the full Hilbert space splits into 17 independent sectors with definite number of fermions. The particle-hole symmetry relates these sectors pairwise. For each state with $n_F$ fermions there is a state with $16-n_F$ fermions and the same energy. Therefore, one needs to study only $n_F\leq8$.

We scale the variables as follows:
\begin{align}
\begin{split}
x^a_i&\rightarrow g^{-1/3} x^a_i,\\
p^a_i&\rightarrow g^{1/3} p^a_i,\\
f^a_\alpha&\rightarrow f^a_\alpha,
\end{split}
\end{align}
so that the only dependency on $g$ is the overall factor $g^{2/3}$ multiplying the Hamiltonian. Therefore, the energy depends trivially on the coupling constant. $g$ will be dropped in the rest of this paper.

Finally, consider parity. Operators transform under parity in the following way:
\begin{align}\label{eq:parity}\begin{split}
p^a_i&\rightarrow-p^a_i,\\
x^a_i&\rightarrow-x^a_i,\\
\psi_\alpha^a&\rightarrow(\psi_P)^{a}_\alpha\equiv\gamma^0_{\alpha\beta}\psi_\beta^a.
\end{split}\end{align}
The Hamiltonian is invariant under the transformation (\ref{eq:parity}). However, the spinor $\psi_P$ does not satisfy the Majorana condition. It follows that states generated by $\psi_P$ are not invariant under the charge conjugation. Therefore, the parity is broken at the level of the Hilbert space. Still, parity is conserved in the bosonic sector.

\section{The cut Fock space method}\label{ch:Fock_space_method}
In this section we discuss a numerical technique, called cut Fock space method, which is used to solve our model. It originates from the variational Tamm-Dancoff method \cite{Dancoff} where one uses a small set of trial states to construct an approximate ground state of a theory. It is also used in the context of quantization on the light cone where the harmonic resolution plays the role of a cutoff \cite{Brodsky}.

The cut Fock space method was already applied with success to other simpler models. Properties of the cut Fock space were studied for one-dimensional quantum mechanics \cite{Trzetrzelewski_spectra}. The technique was used for computations with high precision for the double well potential \cite{Ambrozinski} and multiple wells with periodic boundary conditions \cite{Ambrozinski_cosine}. It was also applied to SYMQM in two dimensions \cite{Trzetrzelewski_susyd2,Korcyl,KorcylN} and finally to four dimensional theory with $SU(2)$ gauge group \cite{Wosiek,Campostrini}.

\subsection{The cut Fock space}
In order to construct the Fock space we introduce creation and annihilation operators, which satisfy the usual commutation rules
\begin{align}\label{eq:bosonic_ca_ops}
a^a_i&=\frac{1}{\sqrt2}(\omega x^a_i+\frac i\omega p^a_i),&a^{a\dagger}_i=\frac{1}{\sqrt 2}(\omega x^a_i-\frac i\omega p^a_i),
\end{align}
\begin{align}
[a^a_i,a^{b\dagger}_j]=\delta^{ab}\delta_{ij}.
\end{align}
At this stage $\omega>0$ is a free parameter and will be used later to improve accuracy of results. The Fock vacuum is defined as usual by
\begin{align}
a^a_i\ket{0}&=0,&f^a_\alpha\ket{0}&=0.
\end{align}
All other states are generated by acting with bosonic and fermionic creation operators on the empty state. However, one can choose only specific combinations of creation operators to generate the space of gauge invariant states. Take a set of operators $\{A_k^a\}$, $k=1,\ldots,n$. The lower index is not related to any symmetry and $A_k^a$ can be any operators. The object
\begin{align}
(A_1\dotsm A_n)\equiv Tr(A_1\dotsm A_n)=A^{a_1}_1\dotsm A^{a_n}_nTr(T^{a_1}\dotsm T^{a_n})
\end{align}
is called a \emph{trace operator}. The number of operators $n$ is referred to as \emph{length} of the trace. For the rest of this paper the round bracket $(\cdot)$ is used for a short notation of the trace. If all operators $A_k$ are bosonic or fermionic creation operators then $(A_1\dotsm A_n)$ is called a \emph{brick}. A product of bricks is called a \emph{composite brick}. It was shown \cite{Trzetrzelewski_trees} that the space of gauge singlets is spanned by states obtained by acting with composite bricks on the Fock vacuum. A state generated by a composite brick has a definite number of bosons $n_B^i$ and fermions $n_F^\alpha$
\begin{align}\label{eq:number_of_particles_operators}
n_B^i&=\sum_a a^{a\dagger}_ia^a_i,&n_F^\alpha&=\sum_a f^{a\dagger}_\alpha f^a_\alpha.
\end{align}
Therefore, we associate the occupation labels $\mathbf n=(n^1_F,n^2_F,n^1_B,n^2_B,n^3_B)$ with a composite brick.

The Hamiltonian conserves $n_F$, so it is convenient to work with subspaces with fixed number of fermions $\mathcal H_{n_F}$. The cut Fock space $\mathcal H_{n_F,N_B}$ is then the space of all states which contain precisely $n_F$ fermions and $n_B\equiv\sum_i n^i_B\leq N_B$. The cutoff can be different for each $n_F$.

The cut Fock space can be decomposed into subspaces with definite occupation numbers
\begin{align}\label{eq:Hilbert_space_decomposition}
\mathcal H_{n_F,N_B}=\bigoplus_\mathbf n\mathcal H_\mathbf n,
\end{align}
where components of $\mathbf n\geq0$ satisfy $\sum_\alpha n^\alpha_F=n_F$ and $\sum_i n^i_B\leq N_B$. A subspace $\mathcal H_{\mathbf n}$ is spanned by all composite bricks with occupation labels $\mathbf n$ acting on the Fock vacuum.

The cut Fock space method will be used to construct matrices for several operators. The angular momentum operators conserve $n_F$ and $n_B$, so the cut matrices of these operators are $(J_i)_{n_F,N_B}:\mathcal H_{n_F,N_B}\rightarrow \mathcal H_{n_F,N_B}$. Eigenvalues of these matrices are exact eigenvalues of $J_i$. 

The Hamiltonian conserves $n_F$ but not $n_B$. The matrix $H_{n_F,N_B}:\mathcal H_{n_F,N_B}\rightarrow \mathcal H_{n_F,N_B}$ has eigenvalues which approximate energy levels of the Hamiltonian in the $N_B\to\infty$ limit.
Finally, matrices of $Q_\alpha$ will be constructed. Supercharges do not conserve $n_F$. In Chapter \ref{ch:supersymmetric_multiplets} we introduce $\mathcal Q_\pm$ which also play role of supercharges. Operators $\mathcal Q_\pm$ decrease the number of fermions by $1$. Therefore, we generate matrices $(\mathcal Q_{\pm})_{n_F,N_B,N_B'}:\mathcal H_{n_F,N_B}\rightarrow \mathcal H_{n_F-1,N_B'}$. In practice the two cutoffs $N_B$ and $N_B'$ are always different.

\subsection{Relations between bricks for $SU(3)$}\label{sec:brick_relations}
The set all composite bricks acting on the empty state is an overcomplete basis of the full Fock space. For optimization reasons it is necessary to have as few basis states as possible. In this section we identify and remove those bricks which can be expressed in terms of other bricks.

The Cayley Hamilton theorem states that a matrix is a root of its characteristic polynomial. Let $M$ be a square traceless matrix of size 3. Then, the theorem implies that
\begin{align}\label{eq:CH}
M^3&=(M)M^2+\frac{1}{2}\left\{(M^2)-(M)^2\right\}M+\frac{1}{6}\left\{(M)^3-3(M^2)(M)+2(M^3)\right\}\mathds{1}_3.
\end{align}
Recall that $(\cdot)$ is a short notation for a trace. This theorem holds if the matrix is operator-valued, i.e. its elements are operators, as long as the matrix elements commute. Multiply the above equation by another operator-valued matrix $O$ and take a trace. Then,
\begin{align}\label{eq:CH_trace}
(M^3O)&=(M)(M^2O)+\frac{1}{2}\left\{(M^2)-(M)^2\right\}(MO)+\frac{1}{6}\left\{(M)^3-3(M^2)(M)+2(M^3)\right\}(O).
\end{align}
It follows that if a brick contains an expression which is repeated three times and at least one more operator, then it can be written in terms of shorter bricks and therefore is redundant. The simplest example of such brick is $(a_1^\dagger a_1^\dagger a_1^\dagger f_2^\dagger)$.

There are other relations for fermionic operators. If $M$ is an operator-valued matrix with anticommuting matrix elements, then the following identities hold \cite{Procesi}:
\begin{align}\begin{split}
(M^2)=(M^4)&=0,\\
M^5&=\frac{1}{3}(M^3)M^2+\frac{1}{3}(M^5),\\
M^6&=0.
\end{split}\end{align}
Another identity for generators of $SU(3)$ reads \cite{Macfarlane}
\begin{align}\label{eq:symmetric_reduction}
T^{\{a}T^bT^{c\}}=\frac{1}{4}\delta^{\{ab}T^{c\}}+\frac{1}{3}(T^{\{a}T^bT^{c\}})\mathds 1_3.
\end{align}
Curly brackets denote symmetrization without additional coefficient $\frac{1}{3!}$. For any six bricks that differ by permutations of three operators, one of them can be eliminated.

There is one more relation \cite{doktorat} concerning $T^a$, namely a product of six $SU(3)$ generators $T^aT^bT^cT^dT^eT^f$ can be expressed in terms of its trace and a linear combination of shorter products multiplied by some tensors. Each of these tensors is a product of traces of products of generators. Each trace has less than six generators inside. It follows that any brick longer than six can be decomposed into shorter bricks and it is therefore redundant. It follows that there is only a finite number of brick in total.

Finally, if a set of composite bricks $\{\mathfrak B_i\}$ acting on the Fock vacuum gives linearly dependent states, i.e.
\begin{align}\label{eq:dependent_paths}
\sum_i\alpha_i\mathfrak B_i\ket{0}&=0,&\alpha_i\neq0,
\end{align}
then the composite bricks themselves are linearly dependent:
\begin{align}\label{eq:dependent_paths2}
\sum_i\alpha_i\mathfrak B_i&=0.
\end{align}

In each sector with given occupation labels $\mathbf n$ one can use the Gauss elimination method to find all relations of the kind (\ref{eq:dependent_paths}). If at least one brick $\mathfrak B_i$ is not composite, then it is redundant by the virtue of (\ref{eq:dependent_paths2}). A single relation (\ref{eq:dependent_paths2}) allows to discard only one brick. The Gauss elimination method allows one to find the complete set of independent bricks. The Gramm matrix required for the Gauss elimination is constructed according to the following part of this section. We found that there are 786 independent bricks.

We proposed a method to eliminate all dependent bricks. Although their number is finite, there is an infinite number of composite bricks. This is obvious, because the Hilbert space is infinite-dimensional. Moreover, composite bricks are linearly dependent, e.g. $(a_1^\dagger f_2^\dagger)(a_1^\dagger a_1^\dagger)=(a_1^\dagger a_1^\dagger)(a_1^\dagger f_2^\dagger)$. This dependency cannot be removed if one wants to take advantage of the following recursive algorithm.

\subsection{Matrix elements}\label{sec:matrix_elements}
In this part we present the algorithm for constructing matrix elements of operators. Although this paper concentrates on the $SU(3)$ gauge group, the algorithm is suited for the general case with $SU(N)$ group.

Let us first introduce some notation. A \emph{composite trace operator} is a product of trace operators. \emph{Length} of the composite trace is the total of lengths of traces in the product. All operators of our interest, i.e. angular momentum, Hamiltonian and supercharges can be expressed by combinations of composite trace operators. We assume that the traces consist only of creation and annihilation operators.

Take an arbitrary gauge invariant operators $A$. The full Hilbert space is decomposed into orthogonal subspaces $\mathcal H_\mathbf n$ and each subspace is spanned by composite bricks with occupation numbers $\mathbf n$ acting on the Fock vacuum. The matrix of operator $A$ can be written in a block form with blocks $A|_{\mathbf n'\mathbf n}:\mathcal H_\mathbf n\rightarrow\mathcal H_{\mathbf n'}$. Dimension $D_\mathbf n$ of $\mathcal H_\mathbf n$ is finite for each $\mathbf n$. The dimension can be determined numerically as well as with an analytic method \cite{Trzetrzelewski_number,doktorat}. $A$ can be expressed as a combination of composite traces. Then block $A|_{\mathbf n'\mathbf n}$ is a relevant combination of products of blocks of trace operators. From now on we assume that $A$ is a trace operator itself, i.e. $A=(A_1\ldots A_m)$.

Let $\hat n^\alpha_F$ be the number of creation operators $f_\alpha^\dagger$ in $A$ minus the number of annihilation operators $f_\alpha$. Let $\hat n^i_B$ be defined in analogous way. Then $\hat{\mathbf n}=(\hat n^1_F,\hat n^2_F,\hat n^1_B,\hat n^2_B,\hat n^3_B)$ is called \emph{creation labels} for $A$. Obviously, if $\mathbf n'\neq\mathbf n+\hat{ \mathbf n}$ then the block $A|_{\mathbf n'\mathbf n}$ vanishes. From now on we consider only $\mathbf n'=\mathbf n+\hat{ \mathbf n}$.

We shall now make some assumptions. First, $A$ contains at least one annihilation operator. If this is not the case, then $A$ is proportional to the identity or contains only creation operators. In the former case, the block $A|_{\mathbf n'\mathbf n}$ is proportional to the identity matrix. If $A$ has only creation operators, then we construct $A|_{\mathbf n'\mathbf n}=\left(A^\dagger|_{\mathbf n\mathbf n'}\right)^\dagger$ and $A^\dagger$ contains only annihilation operators. Secondly, assume that $A=(A_1\ldots A_m)$ annihilates the empty state. Otherwise, a certain cyclic permutation $\tilde A=(A_i\cdots A_mA_1\cdots A_{i-1})$ does annihilate the Fock vacuum (e.g. when $A_{i-1}$ is an annihilation operator). The remainder $A-\tilde A$ is a combination of composite trace operators. A standard example is $(a^{\phantom\dagger}_1a_1^\dagger)=(a_1^\dagger a_1^{\phantom\dagger})+\frac12$. Obviously, $A|_{\hat{\mathbf n}\mathbf 0}$ is a zero matrix. More assumptions can be put on the form of $A$ to improve efficiency (cf. \cite{doktorat}). However, they are not essential for correctness of the algorithm.

Let $\{\mathcal B_i\}$ be the set of all bricks that contain at least one creation operator corresponding to the first nonzero component of $\mathbf n$ (e.g. if $n_F^1\neq0$ it is $f^\dagger_1$ and if $n_F^1=0\neq n_F^2$ then it is $f^\dagger_2$, etc.). The creation labels of $\mathcal B_i$ will be denoted by $\hat{\mathbf n}_i$. Next, remove all bricks for which $\mathcal H_{\mathbf n_i}$, where $\mathbf n_i=\mathbf n-\hat{\mathbf n}_i$, is empty. In particular, all bricks for which at least one component of $\mathbf n_i$ is negative have to be removed. $\mathcal H_{\mathbf n}$ is then spanned by composite bricks acting on Fock vacuum, each containing at least one brick from the set $\{\mathcal B_i\}$. Now, take an orthonormal basis $\ket{e^{\mathbf n_i}_j}$ in the sector $\mathcal H_{\mathbf n_i}$. Vectors $\ket{v_k}=\mathcal B_i\ket{e^{\mathbf n_i}_j}$ span $\mathcal H_\mathbf n$. The index $k$ enumerates all pairs $(i,j)$ on the right hand side. Vectors $\ket{v_k}$ may be not orthogonal and the number of them can be greater than the dimension $D_{\mathbf n}$ of $\mathcal H_\mathbf n$. This will be taken into account later by orthogonalization matrix.

The block $A|_{\mathbf n'\mathbf n}$ is constructed in two steps. First, a block in the overcomplete basis is built:
\begin{align}
(\bar{A}|_{\mathbf n'\mathbf n})_{lk}=\braket{e^{\mathbf n'}_l|A|v_k}=\braket{e^{\mathbf n'}_l|A\mathcal B_i|e^{\mathbf n_i}_j}.
\end{align}
It can be written in a block form:
\begin{align}
\bar{A}|_{\mathbf n'\mathbf n}=
\left(\begin{array}{c}
(A\mathcal B_1)|_{\mathbf n'\mathbf n_1}\\
\vdots\\
(A\mathcal B_q)|_{\mathbf n'\mathbf n_q}
\end{array}\right).
\label{eq:block_matrix}
\end{align}
Each of the sub-blocks is constructed as
\begin{align}
(A\mathcal B_i)|_{\mathbf n'\mathbf n_i}&=\mathcal B_i|_{\mathbf n',\mathbf n'-\hat{\mathbf n}_i}\times A|_{\mathbf n'-\hat{\mathbf n}_i,\mathbf n_i}+[A,\mathcal B_i]|_{\mathbf n'\mathbf n_i}
\end{align}
and each of the blocks on the right hand side is built recursively. Once the block $\bar{A}|_{\mathbf n'\mathbf n}$ is constructed, we orthogonalize the basis of $\mathcal H_\mathbf n$. Let $S_\mathbf n$ be the Gramm matrix in the sector $\mathcal H_\mathbf n$:
\begin{align}
S_{\mathbf n}&=
\left(\begin{array}{ccc}
(\mathcal B_1^\dagger \mathcal B_1)|_{\mathbf n_1\mathbf n_1}&\ldots& (\mathcal B_1^\dagger \mathcal B_q)|_{\mathbf n_1\mathbf n_q}\\
\vdots&\ddots&\vdots\\
(\mathcal B_q^\dagger \mathcal B_1)|_{\mathbf n_q\mathbf n_1}&\ldots& (\mathcal B_q^\dagger \mathcal B_q)|_{\mathbf n_q\mathbf n_q}\\
\end{array}\right).
\label{eq:gramm_matrix}
\end{align}
As before, each sub-block is constructed as
\begin{align}
(\mathcal B_i^\dagger \mathcal B_j)|_{\mathbf n_i\mathbf n_j}&=\mathcal B_j|_{\mathbf n_i\mathbf n}\times \mathcal B_i^\dagger|_{\mathbf n\mathbf n_j}+[\mathcal B_i^\dagger, \mathcal B_j]|_{\mathbf n_i\mathbf n_j}.
\end{align}

Matrix $S_\mathbf n$ has exactly $D_\mathbf n$ nonzero eigenvalues $\lambda_l$. Corresponding eigenvectors $w^l_k$ are used to produce the basis of $\mathcal H_\mathbf n$. More precisely, the orthonormal basis is given by
\begin{align}
\ket{e_l^{\mathbf n}}=\frac{1}{\sqrt{\lambda_l}}\sum_k w^l_k\ket{v_k}.
\end{align}
Finally, we construct the orthogonalization matrix $R_\mathbf n$ by setting its matrix elements to $(R_\mathbf n)_{kl}=\frac{1}{\sqrt\lambda_l}w^l_k$. Then, $A|_{\mathbf n'\mathbf n}$ is the product of the block in overcomplete basis and the orthogonalization matrix:
\begin{align}
(\bar{A}|_{\mathbf n'\mathbf n}R_\mathbf n)_{ij}=\sum_k(\bar{A}|_{\mathbf n'\mathbf n})_{ik}(R_\mathbf n)_{kj}=\sum_k\braket{e^{\mathbf n'}_i|A|v_k}\frac{1}{\sqrt\lambda_j}w^j_k=\braket{e^{\mathbf n'}_i|A|e_j^{\mathbf n}}.
\end{align}

\subsection{Diagonalization}\label{sec:diagonalization}

Some remarks concerning diagonalization of matrices are in place. Assume that we constructed matrices of the Hamiltonian $H$, square of total angular momentum $J^2$ and the third component of angular momentum $J_3$ in a sector with $n_F$ fermions and cutoff $N_B$. In order to obtain energies one can diagonalize $H$ and then eventually act with $J^2$ and $J_3$ on eigenvectors to check what are their quantum numbers. This procedure is however ineffective.

Operators $J^2$ and $J_3$ conserve the number of bosons $n_B$ while $H$ does not. Therefore, matrices of angular momentum decompose into smaller matrices on subspaces with fixed $n_B$:

\begin{align}
J^2_{N_B}&=\left(
\begin{array}{ccc}
J^2_{n_B=1}&&\multirow{2}{*}{\LARGE 0}\\
\multirow{2}{*}{\LARGE 0}&\ddots&\\
&&J^2_{n_B=N_B}
\end{array}\right),&
(J_3)_{N_B}&=\left(
\begin{array}{ccc}
(J_3)_{n_B=1}&&\multirow{2}{*}{\LARGE 0}\\
\multirow{2}{*}{\LARGE 0}&\ddots&\\
&&(J_3)_{n_B=N_B}
\end{array}\right).
\end{align}
Index $n_F$ is dropped for shorter notation.

We diagonalize $J^2_{n_B}$ for each $n_B$. Since these matrices are smaller, it is significantly faster to diagonalize them than $H$. For each value of angular momentum $j$ we construct a projection matrix $P^{n_B}_j$. It maps $\mathcal H_{n_B}$ onto a subspace $\mathcal H_{n_B,j}$ corresponding to given $j$. Then $J_3$ on these small subspaces is
\begin{align}
(J_3)_{n_B,j}=P^\dagger_{n_Bj}(J_3)_{n_B}P^{\phantom\dagger}_{n_Bj}.
\end{align}
These matrices are yet smaller and can be diagonalized for each $n_B$ and $j$ separately. Projection matrices $P_{n_B,j,m}$ from $\mathcal H_{n_B,j}$ to $\mathcal H_{n_B,j,m}$ are given by eigenvectors of $(J_3)_{n_B,j}$. Then the projection matrices from $\mathcal H_{n_B}$ to $\mathcal H_{n_B,j,m}$ are $\mathcal P_{n_B,j,m}=P_{n_B,j,m}P_{n_B,j}$. We construct a transition matrix
\begin{align}
\mathcal P_{N_B,j,m}&=\left(
\begin{array}{ccc}
\mathcal P_{n_B=1,j,m}&&\multirow{2}{*}{\LARGE 0}\\
\multirow{2}{*}{\LARGE 0}&\ddots&\\
&&\mathcal P_{n_B=N_B,j,m}
\end{array}\right).
\end{align}
Finally, the matrix of Hamiltonian in a channel with given $(j,m)$ is a product $H_{N_B,j,m}=\mathcal P_{N_B,j,m}^\dagger H_{N_B}\mathcal P^{\phantom\dagger}_{N_B,j,m}$. It is much smaller than the initial full matrix $H_{N_B}$ and thus diagonalization is faster.

Energies in the cut Fock space method are approximated from above. The parameter $\omega$ (cf. (\ref{eq:bosonic_ca_ops})) is not yet fixed. It can be used to minimize the lowest eigenvalue of the Hamiltonian and thus to improve approximation of the ground energy. Matrix representations of the angular momentum operators are independent of $\omega$. Therefore, projection matrices $\mathcal P_{N_B,j,m}$ can be constructed without fixing $\omega$. Let us now construct matrices of three terms of the Hamiltonian $\bar H_k,\ \bar H_V,\ \bar H_F$ with $\bar\omega=1$ and project them onto the subspace with given $j$ and $m$. Then the matrix of the full Hamiltonian is
\begin{align}\label{eq:omega_hamiltonian}
H_{N_B,j,m}(\omega)=\omega^2 (\bar H_K)_{N_B,j,m}+\omega^{-4}(\bar H_V)_{N_B,j,m}+\omega^{-1}(\bar H_F)_{N_B,j,m}.
\end{align}
Matrix of the Hamiltonian in the subspace with given angular momentum is relatively small and minimizing its smallest eigenvalue with respect to $\omega$ is fast.

We come to discussing the effect of including $\omega$. Results concerning the energies are given in Tab. \ref{tab:energies_omega_min}. For $(n_F,j)=(0,0)$ the smallest eigenvalue with cutoff $N_B$ minimized with respect to $\omega$ is smaller then the lowest eigenvalues for cutoff $N_B+2$ with $\omega=1$. That is, including the parameter $\omega$ effectively increases the cutoff by $2$. The effect is similar for $(n_F,j)=(2,1)$. In the singlet channel for six fermions inclusion of $\omega$ effectively rises $N_B$ by almost $1$.

\newcolumntype{d}[1]{D{.}{.}{#1}}
\def\arraystretch{1.3}
\begin{table}\centering
\begin{tabular}{|c||d{-1}|d{-1}||d{-1}|d{-1}||d{-1}|d{-1}|}
\hline
\multirow{3}{*}{$N_B$}&\multicolumn{6}{c|}{lowest eigenvalue in sector with quantum numbers $(n_F,j)$}\\
\cline{2-7}&\multicolumn{2}{c||}{$(0,0)$}&\multicolumn{2}{c||}{$(2,1)$}&\multicolumn{2}{c|}{$(6,0)$}\\
\cline{2-7}&\multicolumn{1}{c|}{$\omega=1$}&\multicolumn{1}{c||}{$\omega=\omega_{min}$}&\multicolumn{1}{c|}{$\omega=1$}&\multicolumn{1}{c||}{$\omega=\omega_{min}$}
&\multicolumn{1}{c|}{$\omega=1$}&\multicolumn{1}{c|}{$\omega=\omega_{min}$}\\\hline
 0&15       &12.992    &       &           &15        &12.99\\
 1&15       &12.992    &17     &14.41      &9.06      &7.83\\
 2&13.327   &12.952    &14.35  &12.15      &6.12      &5.10\\
 3&13.327   &12.952    &13.03  &11.47      &4.59      &3.74\\
 4&12.882  &12.632    &12.05  &11.25      &          &\\
 5&12.882  &12.632    &11.54  &11.04      &          &\\
 6&12.712  &12.620    &11.20  &10.79      &          &\\
 7&12.712  &12.620    &10.97  &10.64      &          &\\
 8&12.634  &12.591    &       &           &          &\\
 9&12.634  &12.591    &       &           &          &\\
10&12.604  &12.589    &       &           &          &\\\hline
\end{tabular}
\caption{The lowest energies in selected sectors. In each case the energy is given with keeping $\omega=1$ and minimizing the energy with respect to $\omega$.}
\label{tab:energies_omega_min}
\end{table}

\def\arraystretch{1.2}
\begin{table}\centering
\begin{tabular}{|c|d{-1}|d{-1}|d{-1}|d{-1}|d{-1}|d{-1}|d{-1}|d{-1}|d{-1}|}
\hline
\diaghead{\theadfont Diag Colu}{$j$}{$n_F$}&
0&1&2&3&4&5&6&7&8\\
\hline
$0$             &1.22&&1.21&&1.15&&1.15&&1.14\\
$1/2$   &&1.31&&1.23&&1.15&&1.16&\\
$1$             &1.25&&1.22&&1.15&&1.15&&1.15\\
$3/2$   &&1.27&&1.22&&1.15&&1.15&\\
$2$             &1.25&&1.20&&1.15&&1.16&&1.15\\\hline
\end{tabular}
\caption{Optimal values of $\omega$ in each $(n_F,j)$ channel. The value depends on the cutoff and is given for the maximal available cutoff.}
\label{tab:omega_min}
\end{table}
\def\arraystretch{1}

In Tab. \ref{tab:omega_min} optimal values of $\omega$ for different channels are given. All values of $\omega$ are similar. They are slightly smaller for $n_F\geq 4$. This means that the wavefunctions are in general wider for high $n_F$. This may be a sign of continuous spectrum in these sectors because states corresponding to the continuous spectrum are not localized.

\section{The spectrum}\label{ch:spectrum}

Spectrum of a Hamiltonian is a basic characteristic of a model. In particular, it is interesting to identify the type of the spectrum, whether it is continuous or discrete. The nature of the spectrum can be identified from the behavior of energy levels for growing cutoff \cite{Trzetrzelewski1}.

In order to obtain the spectrum we constructed matrices with sizes up to 36000. Computations were performed on a supercomputer Deszno located in the Institute of Physics at Jagiellonian University. 96 cores were used in parallel with the OpenMP interface using up to 256 GB shared memory.

\subsection{Spectra in sectors with given number of fermions}
\begin{figure}[!h]\centering
\begin{subfigure}[b]{0.3\textwidth}\includegraphics[width=\textwidth]{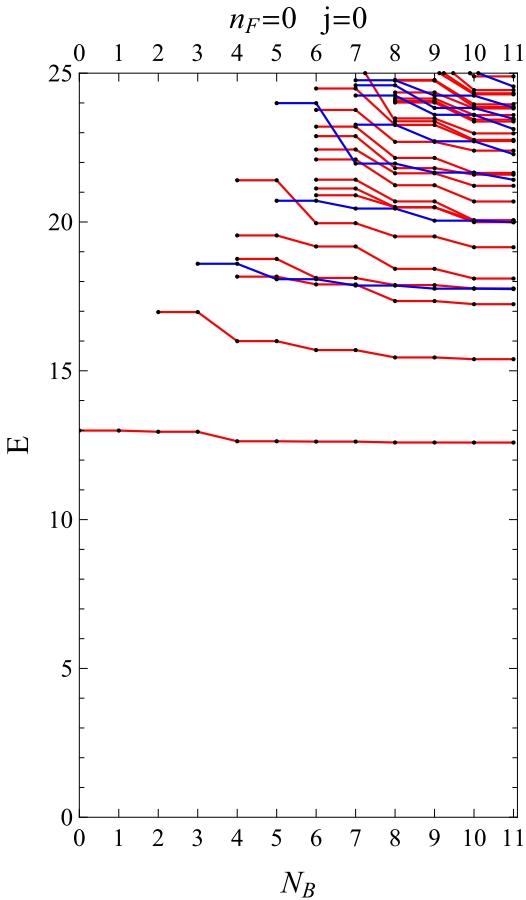}\end{subfigure}
\begin{subfigure}[b]{0.3\textwidth}\includegraphics[width=\textwidth]{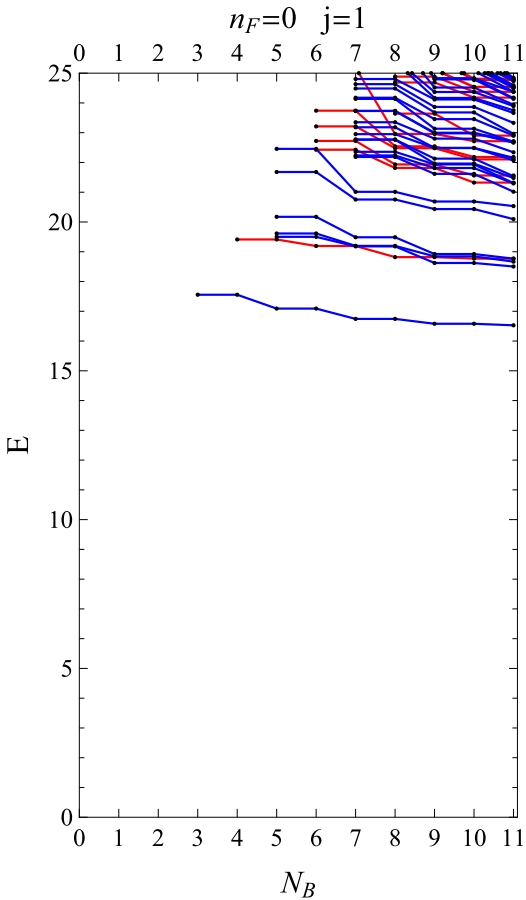}\end{subfigure}
\begin{subfigure}[b]{0.3\textwidth}\includegraphics[width=\textwidth]{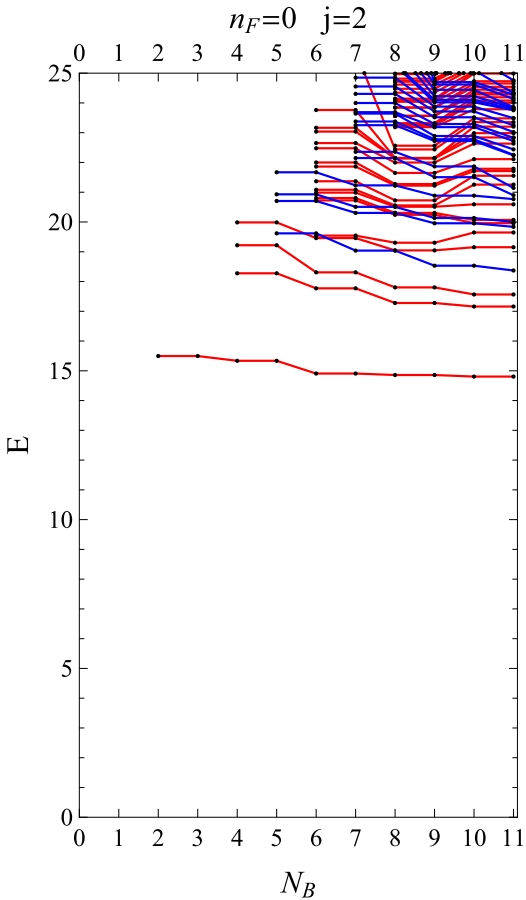}\end{subfigure}
\caption{Spectrum of the $SU(3)$ theory in the bosonic sector. The maximal cutoff is $N_B=11$. Energies marked with red correspond to states with positive parity and blue to negative parity.}
\label{fig:su3_spectrum_nf0}
\end{figure}

We first comment on the 'stepwise' convergence of eigenvalues for $n+F=0$ (cf. Fig. \ref{fig:su3_spectrum_nf0}). In the bosonic channel parity is conserved. Creation operators $a_i^\dagger$ have negative parity, so states with odd $n_B$ (see eq. (\ref{eq:number_of_particles_operators})) have negative parity. When $N_B$ raises by $1$, it changes e.g. from an even to odd value, so only the basis of odd states grows. Therefore, eigenvalues corresponding to even states remain unchanged. Similarly, energy levels of odd states do not change when $N_B$ changes from an odd to even value.

\begin{figure}[!h]\centering
\begin{subfigure}[b]{0.3\textwidth}\includegraphics[width=\textwidth]{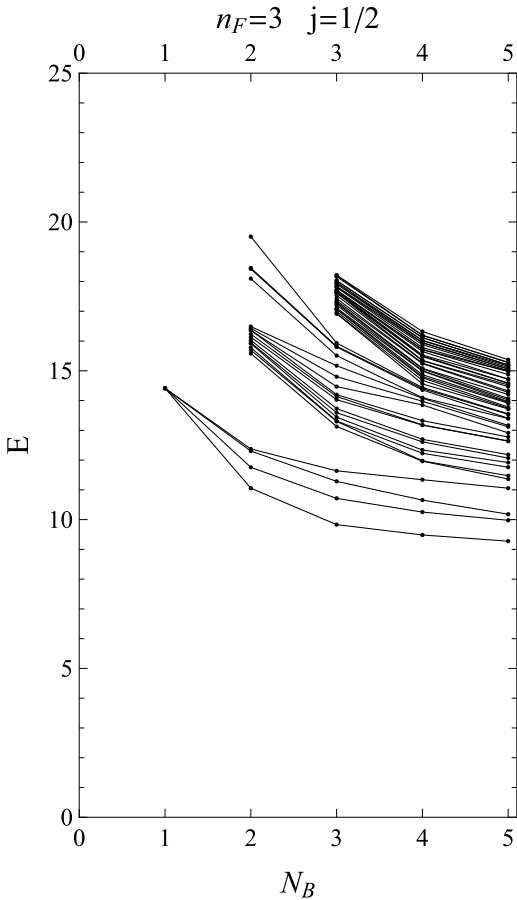}\end{subfigure}
\begin{subfigure}[b]{0.3\textwidth}\includegraphics[width=\textwidth]{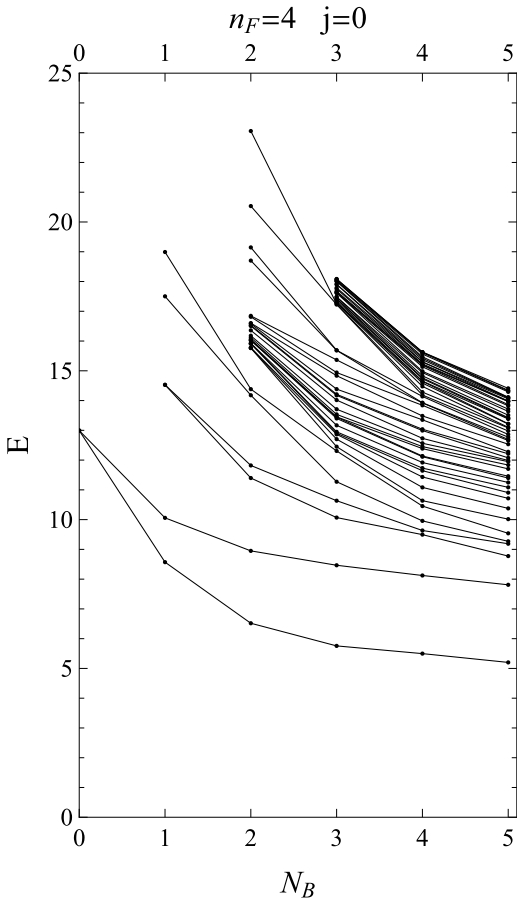}\end{subfigure}
\begin{subfigure}[b]{0.3\textwidth}\includegraphics[width=\textwidth]{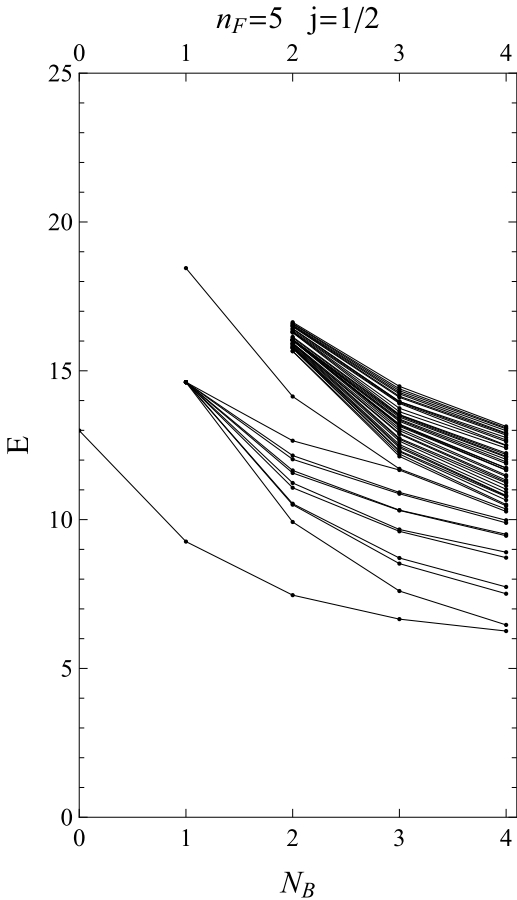}\end{subfigure}
\caption{Spectrum of the $SU(3)$ theory with $n_F=3,4,5$ and the lowest angular momentum.}
\label{fig:su3_spectrum_nf_small}
\end{figure}

For $n_F=0$ all energies converge fast with growing cutoff. This behavior is typical for discrete spectrum. The fact that the spectrum is discrete is contrary to what one may naively expect. Indeed, if the three matrices $x_i=x_i^aT^a$ commute then the potential vanishes: $V\equiv-\sum_{i<j}Tr([x_i,x_j]^2)=0$. The region where the potential vanishes forms a vector subspace in the configuration space which we call flat valleys. In such situation the spectrum is usually continuous. On the other hand, the potential becomes steeper in the transverse directions as one moves deeper inside the valleys. The transverse oscillations cost more energy and thus the effective potential inside the flat valleys, integrated over transverse degrees of freedom, grows. Therefore, the energy eigenstates are localized and the spectrum is discrete. These properties are shared by some simpler systems, see e.g. \cite{Simon}. Similar behavior is observed in sectors with $n_F=1,\dots,4$ fermions, although the ground energy decreases as $n_F$ grows (cf. Fig \ref{fig:su3_spectrum_nf_small}). This is different from the case of $SU(2)$ where there is continuous spectrum for $n_F=2,3,4$ \citation{Campostrini}.

\begin{figure}[!h]\centering
\begin{subfigure}[b]{0.3\textwidth}\includegraphics[width=\textwidth]{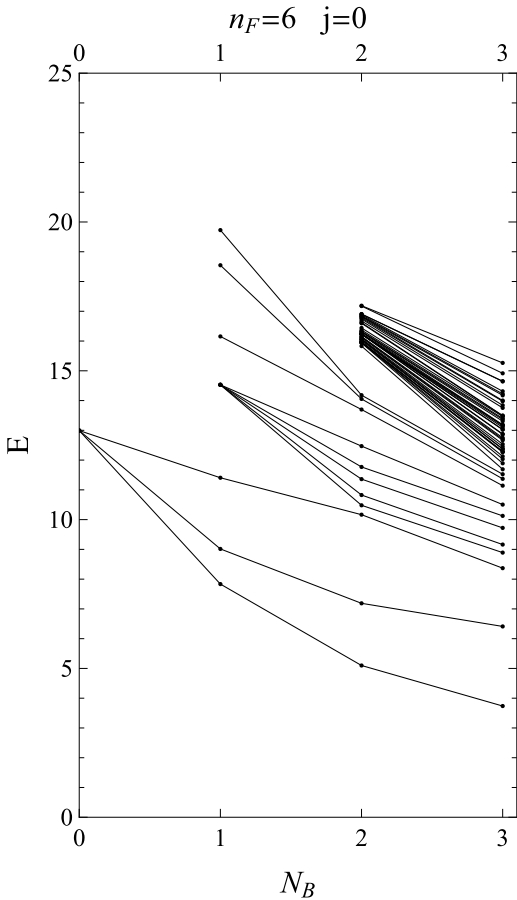}\end{subfigure}
\begin{subfigure}[b]{0.3\textwidth}\includegraphics[width=\textwidth]{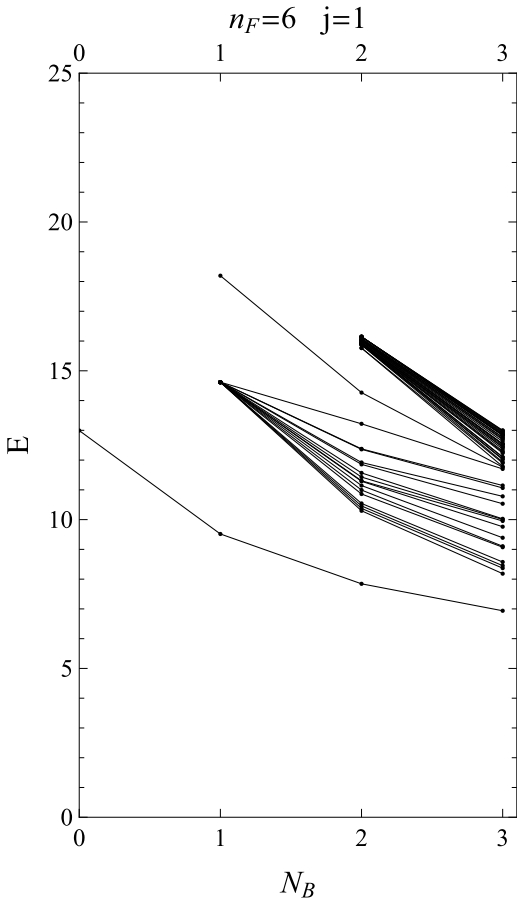}\end{subfigure}
\begin{subfigure}[b]{0.3\textwidth}\includegraphics[width=\textwidth]{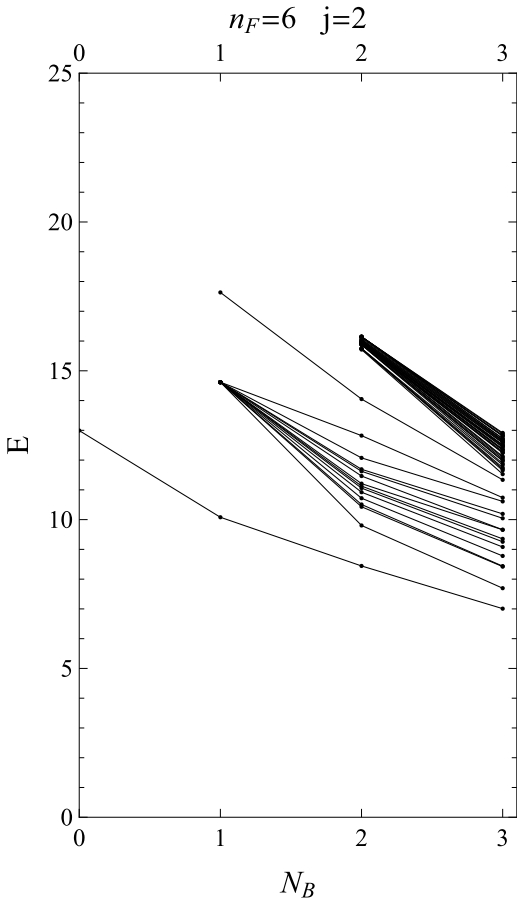}\end{subfigure}
\caption{Spectrum of the $SU(3)$ theory with $n_F=6$.}
\label{fig:su3_spectrum_nf6}
\end{figure}

\begin{figure}[!h]\centering
\begin{subfigure}[b]{0.3\textwidth}\includegraphics[width=\textwidth]{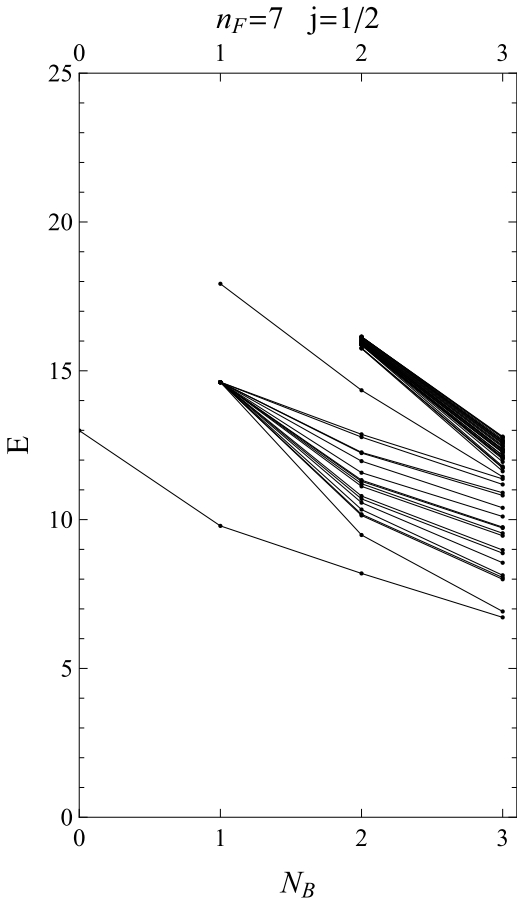}\end{subfigure}
\begin{subfigure}[b]{0.3\textwidth}\includegraphics[width=\textwidth]{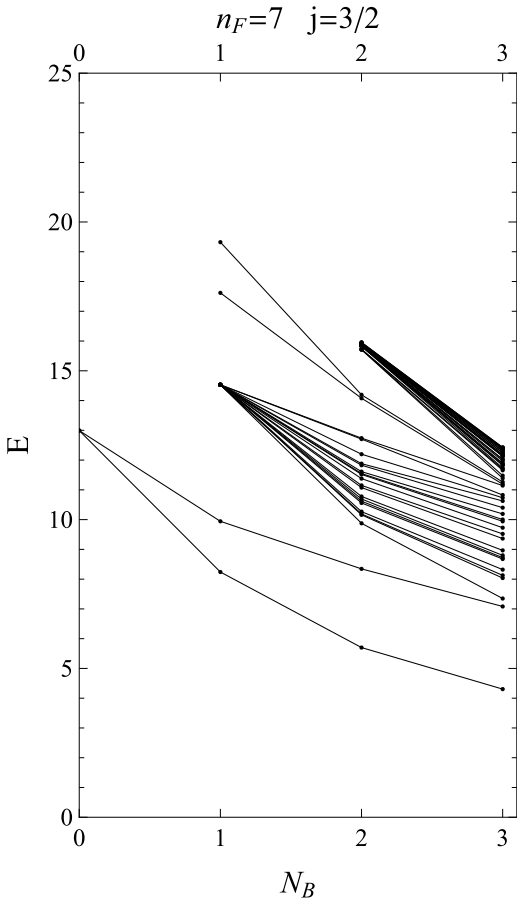}\end{subfigure}
\begin{subfigure}[b]{0.3\textwidth}\includegraphics[width=\textwidth]{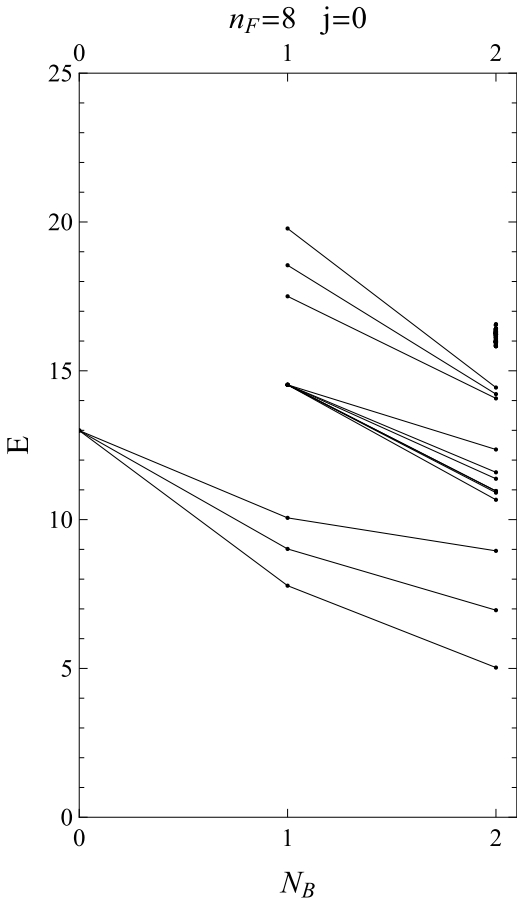}\end{subfigure}
\caption{Spectrum of the $SU(3)$ theory with $n_F=7,8$.}
\label{fig:su3_spectrum_nf_large}
\end{figure}

An alternative possibility is the continuous spectrum. If it is present, then the eigenvalues fall slowly to zero (in the case of no mass gap). This behavior was confirmed for all systems with continuous spectrum studied so far with the cut Fock space method \cite{Trzetrzelewski_spectra,Trzetrzelewski1,Korcyl,Campostrini}. In our results this effect is most manifest for $n_F=6$ and $j=0$ (Fig. \ref{fig:su3_spectrum_nf6}). The continuum spectrum arises because inside the flat valley the contributions from interactions of bosonic and fermionic degrees of freedom with the potential have different sign. They cancel exactly causing the effective potential to vanish. The flat valleys are open. Similar signs of continuous spectrum are also seen for $n_F=7,8$ (Fig. \ref{fig:su3_spectrum_nf_large}).


All these effects, i.e. nonvanishing effective potential in a flat valley, cancelation between bosons and fermions and finally discrete spectrum inside continuous one were already observed for simpler supersymmetric systems \cite{Simon,Nicolai,Korcyl_effective}.

The Yang-Mills quantum mechanics without fermions was considered as a zeroth order approximation to pure gauge $SU(3)$ theory in the small volume expansion in \cite{Ziemann_phd,Weisz}. Their results are obtained numerically by a variational technique, which is essentially the same as our method (but restricted to $n_F=0$). In Table \ref{tab:Weisz_comparison} we present comparison of the lowest energies in several channels with given angular momentum and parity. The results are consistent. The relative difference in most cases does not exceed 4\%. Naturally, the results differ less in cases where both methods are more precise, i.e. for lowest energies.

\begin{table}[h]\centering
\begin{tabular}{|c|c|c|c|}
\hline
sector&cut Fock space&Rayleigh-Ritz&relative difference\\\hline
\multirow{3}{*}{$0^{+}$}&12.5889&12.5887&0.0016\%\\&15.39&15.38&0.07\%\\&17.24&17.23&0.06\%\\\hline
\multirow{1}{*}{$0^{-}$}&17.75&17.8&-0.28\%\\\hline
\multirow{2}{*}{$1^{+}$}&18.77&18&4.1\%\\&22.4&>20&\\\hline
\multirow{2}{*}{$1^{-}$}&16.52&17.05&-3.2\%\\&18.5&23&-24\%\\\hline
\multirow{2}{*}{$2^{+}$}&14.806&14.854&-3.2\%\\&17.159&17.26&-0.59\%\\\hline
\multirow{1}{*}{$2^{-}$}&18.37&>20&\\\hline
$3^{-}$&16.10&16.5&-2.5\%\\\hline
$4^{+}$&17.3&18&-4\%\\\hline
\end{tabular}
\caption{Comparison of lowest energies in the bosonic sector for the $SU(3)$ case obtained by our method and in \cite{Weisz}. Since both approaches are variational, lower values of energies give better approximation to energies in the continuum limit. In the $0^+$ sector results are perfectly consistent -- eigenvalues differ only at the last digit which was given in \cite{Weisz}. In other channels our results are usually slightly more accurate.}
\label{tab:Weisz_comparison}
\end{table}

\subsection{Scaling relations}
There are interesting scaling relations for eigenvalues corresponding to the continuous spectrum. They originate from dispersion relations for energies. In \cite{Trzetrzelewski_spectra} the free Hamiltonian in one dimensional quantum mechanics was studied with the cut Fock space method. For this system the standard dispersion relation $E(p)=\frac{1}{2}p^2$ holds. It was shown that in the infinite $N_B$ limit momentum is related to the label of energy level by $p=\frac{\pi}{\sqrt{2N_B}}(k+1)$. It follows that the energies satisfy
\begin{align}\label{eq:continuous_scaling}
E_k&\approx\frac{\pi^2(k+1)^2}{4N_B}&k=0,1,\ldots
\end{align}
In the continuum limit the energies fill densely the positive real axis. In \cite{Campostrini,Kotanski} it was argued and checked numerically that the scaling of energies are similar for the continuous spectrum in SYMQM. Although the relation (\ref{eq:continuous_scaling}) does not hold precisely, it was shown that $N_BE_k$ converge to nonzero constants.

\begin{figure}[t]\centering
\includegraphics[width=.8\textwidth]{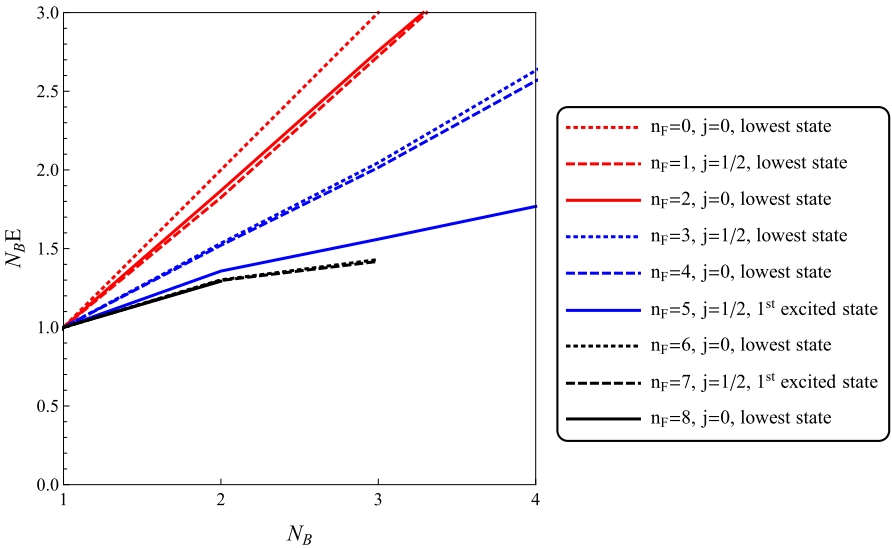}
\caption{Scaled energy $N_BE$ for $SU(3)$ in sectors with $n_F=3,\ldots,8$. All energies are divided by $E$ for $N_B=1$. In each channel the lowest $j$ was picked, i.e. $j=0$ for even $n_F$ and $j=1/2$ for odd $n_F$. In most sectors the lowest energy state was picked. For $n_F=5,7$ the first excited state was chosen because these seem to be better candidates for states in the continuum spectrum. For states corresponding to the continuum spectrum the scaled energy should be flat.}
\label{fig:su3_scaling}
\end{figure}

Corresponding results based on our data are presented in Fig. \ref{fig:su3_scaling}. The scaled energies are given for all $n_F$ in channels with the lowest angular momentum (i.e. $j=0$ for even $n_F$ and $j=1/2$ for odd $n_F$). For $n_F=0,1,2,3,4,6,8$ we picked the lowest energy. For $n_F=5,7$ we chose the first excited energy, because those seem to be better candidates for the continuous spectrum (cf. Figs. \ref{fig:su3_spectrum_nf_small}, \ref{fig:su3_spectrum_nf_large}). Scaled energies corresponding to the continuous spectrum should be flat for large $N_B$. All energies are divided by $E(N_B=1)$ so that the overall scale in each channel is removed. From Fig. \ref{fig:su3_scaling} one can see that the scaled energies flatten as the number of fermions grows.

Concluding, the spectrum of the $SU(3)$ theory has discrete and continuous part. In contrary to the $SU(2)$ case, for $SU(3)$ the spectrum is discrete for two fermions. The continuous spectrum is moved to channels with higher $n_F$ because more fermions are needed to have supersymmetric cancelations between fermionic and bosonic degrees of freedom. Only then the effective potential in the flat valleys vanishes. Our results indicate that this occurs first for $n_F=6$. However, the case of $n_F=5$ is not very clear and the continuous spectrum can be present already in this sector.

\section{Supersymmetry multiplets}\label{ch:supersymmetric_multiplets}
Having computed spectrum of the model, we are now interested in its supersymmetric properties. It is known \cite{Wess} that energy eigenstates of a supersymmetric model form supersymmery multiplets (or supermultiplets).

A powerful method to identify supermultiplets in an organized way is to use supersymmetry fractions \cite{Campostrini}. One can check whether two states are elements of the same multiplet by acting with a supercharge on one of them and analyzing the overlap with the other state. A supersymmetry fraction is essentially such an overlap.

In this section the parameter $\omega$ (cf. Chapter \ref{ch:Fock_space_method}) is not included, i.e. here $\omega=1$.

\subsection{Supersymmetry multiplets in the continuum limit}
The four supersymmetry generators $Q_\alpha$ introduced in (\ref{eq:supercharges}) satisfy anticommutation relations (\ref{eq:anticommutation_relations}) which play a central role in supersymmetric theories. They are however not the only operators that compose the supersymmetric algebra. Moreover, they are somewhat inconvenient because they do not form a spin multiplet. For this reasons we introduce operators $\mathcal Q_\alpha$ ($\alpha=\pm$):
\begin{align}
\mathcal Q_+^\dagger&=-Q_4,&\mathcal Q_+&=Q_1,\nonumber\\
\mathcal Q_-^\dagger&=Q_3&\mathcal Q_-&=Q_2.
\end{align}
Then, $\mathcal Q^\dagger_\alpha$ is spin doublet. Operators $\mathcal Q_\pm^\dagger$ carry magnetic number $m=\pm1/2$. Both $\mathcal Q^\dagger_\pm$ raise the fermionic number $n_F$ by $1$. Their conjugates also form a spin doublet and have opposite quantum numbers. The new supersymmetry generators fulfill the following commutation relations:
\begin{align}\label{eq:anticommutations_modified}
\begin{split}
\{\mathcal Q_\alpha,\mathcal Q_{\beta}\}=\{\mathcal Q_\alpha^\dagger,\mathcal Q_{\beta}^\dagger\}&=0,\\
\{\mathcal Q_\alpha,\mathcal Q_{\beta}^\dagger\}&=2\delta_{\alpha\beta}H,\\
[H,\mathcal Q_\alpha]=[H,\mathcal Q_\alpha^\dagger]&=0.
\end{split}
\end{align}

An irreducible representation of the supersymmetry algebra has a unique Clifford vacuum (or vacuum of the supermultiplet) $\ket{\Omega}$ which satisfies $\mathcal Q_\alpha\ket{\Omega}=0$ for $\alpha=\pm$. However, it is more convenient to work with reducible representations of supersymmetric algebra which contain $SO(3)$ multiplets. Such supermultiplet has several Clifford vacua which compose an angular momentum multiplet $\ket{\Omega,m}$. In what follows we assume that $\ket{\Omega,m}$ has $n_F$ fermions, total angular momentum $j$ and energy $E$.

If $E=0$ then $\mathcal Q^\dagger_\alpha\ket{\Omega,m}$ vanishes and $\ket{\Omega,m}$ is a supersymmetry singlet. If this is not the case, then by repeatedly acting with operators $\mathcal Q^\dagger_\alpha$ on $\ket{\Omega,m}$ one can construct other angular momentum multiplets $\ket{\Omega_-,m},\ \ket{\Omega_+,m},\ \ket{\Omega_{0},m}$. Their quantum numbers are respectively $(n_F+1,j-\frac{1}{2},E),\ (n_F+1,j+\frac{1}{2},E),\ (n_F+2,j,E)$. If $j=0$ then $\ket{\Omega_-,m}=0$. Otherwise none of these states vanishes. Together with the vacuum $\ket{\Omega,m}$ they form a \emph{supersymmetric multiplet}. A supersymmetry multiplet is closed under the action of supercharges. That is, $\mathcal Q_\alpha$ and $\mathcal Q^\dagger_\alpha$ acting on an element of a supermultiplet give either zero or a combination of other members of the same supermultiplet.

The full supersymmetry multiplet forms a diamond in the $(n_F,j)$ plane (cf. Fig. \ref{fig:supermultiplet}). In the case $j=0$ the state $\ket{\Omega_-,m}$ vanishes and the lower node $(n_F+1,j-\frac{1}{2})$ in Fig. \ref{fig:supermultiplet} is not present. A supermultiplet contains $4j+2$ bosonic states (i.e. states which contain even number of fermions) and the same number of fermionic states.

\newlength{\vdiam}
\newlength{\hdiam}
\setlength{\vdiam}{1cm}
\setlength{\hdiam}{3cm}
\begin{figure}
\begin{center}
\begin{tikzpicture}
\node[draw,ellipse, minimum width = 3cm](swest){$n_F,j$};
\node[above right = \vdiam and \hdiam of swest,draw,ellipse, minimum width = 3cm](snorth){$n_F+1,j+\frac{1}{2}$};
\node[below right = \vdiam and \hdiam of swest,draw,ellipse, minimum width = 3cm](south){$n_F+1,j-\frac{1}{2}$};
\node[above right = \vdiam and \hdiam of south,draw,ellipse, minimum width = 3cm](seast){$n_F+2,j$};
\draw[-] (swest) -- (snorth) node [midway,fill=white,draw] {$q=j+1$};
\draw[-] (swest) -- (south) node [midway,fill=white,draw] {$q=j$};
\draw[-] (south) -- (seast) node [midway,fill=white,draw] {$q=j$};
\draw[-] (snorth) -- (seast) node [midway,fill=white,draw] {$q=j+1$};
\end{tikzpicture}
\end{center}
\caption{Structure of a supermultiplet together with supersymmetry fractions. For $j=0$ the bottom vertex and the two lower links are absent.}
\label{fig:supermultiplet}
\end{figure}
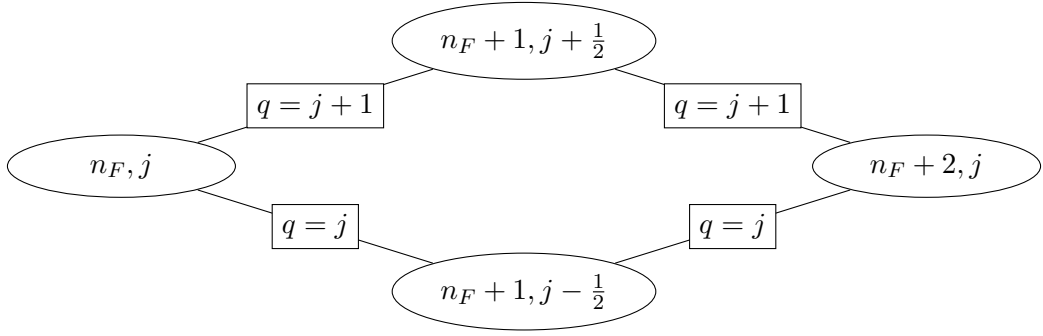

\subsection{Supersymmetry fractions}
A useful tool to analyze the supermultiplets are supersymmetry fractions. They serve two purposes. First is to study breaking of the supersymmetry for finite cutoff. Second is to identify supermultiplets. Consider two angular momentum multiplets $\ket{n_F,j,m,E}$ and $\ket{n_F+1,j',m',E'}$. Two states $\mathcal Q^\dagger_\alpha\ket{n_F,j,m,E}$ and $\ket{n_F+1,j',m',E'}$ have nonzero overlap only if they are in the same supermultiplet and $m'=m+\alpha$. A supersymmetry fraction measures this overlap. It is defined by
\begin{align}
q_{n_F}\left(j',E'|j,E\right)&=\frac{1}{4E}\sum_{mm'\alpha}\left|\braket{n_F+1,j',m',E'|\mathcal Q_\alpha^\dagger|n_F,j,m,E}\right|^2.
\end{align}

Values of the supersymmetry fractions within a supermultiplet are known \cite{doktorat}:
\begin{align}\label{eq:supersummetry_fractions_exact}\begin{split}
q_{n_F}\left(j+\frac{1}{2},E\Big|j,E\right)=q_{n_F+1}\left(j,E\Big|j+\frac{1}{2},E\right)&=j+1,\\
q_{n_F}\left(j-\frac{1}{2},E\Big|j,E\right)=q_{n_F+1}\left(j,E\Big|j-\frac{1}{2},E\right)&=j.
\end{split}\end{align}
The supersymmetry fractions are denoted by rectangles in Fig. \ref{fig:supermultiplet}. The supermultiplet is closed under the action of $\mathcal Q^\dagger_\alpha$ so other fractions vanish.
\begin{figure}[t]\centering
\includegraphics[width=.9\textwidth]{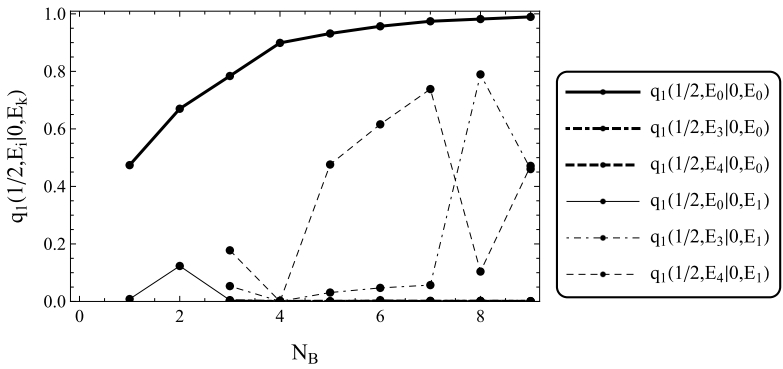}
\caption{Supersymmetry fractions as a function of cutoff. $N_B$ is the cutoff in the sector $n_F=1$. The cutoff for $n_F=0$ is higher by $2$.}
\label{fig:fractions_convergence}
\end{figure}

\subsection{Finite cutoff effects}

For a finite cutoff the supersymmetry is broken. 
Consider a supersymmetry fraction $q_{n_F}(j\pm\frac{1}{2},E_i|j,E_k)$. For large cutoff it is expected that the fraction is zero unless corresponding states, labeled with $i$ and $k$, are in the same supermultiplet. In particular, energies $E_i$ and $E_k$ should be almost equal if the fraction is nonzero. In contrast, for a small cutoff eigenstates of the cut Hamiltonian do not give a good approximation to exact bound states. Then, the supersymmetry fractions $q_{n_F}(j\pm\frac{1}{2},E_i|j,E_k)$ are small but nonzero for many different pairs $(i,k)$.

Analysis of supersymmetry fractions for growing cutoff provides us the information at which value of $N_B$ the supersymmetry is approximately restored. In Fig. \ref{fig:fractions_convergence} dependence of supersymmetry fractions $q_1(\frac{1}{2},E_i|0,E_k)$ on the cutoff is shown for selected $i$ and $k$. All states with $(n_F,j)=(0,0)$ are Clifford vacua, so the exact values of these supersymmetry fractions are either $0$ or $1$. For the ground state in the bosonic sector there is only one fraction, $q_1(\frac{1}{2},E_0|0,E_0)$, which is significant. It approaches $1$ fast and one can say that the supersymmetry between the two corresponding states is essentially restored for $N_B\approx7$. In general higher cutoffs are required to regain supersymmetry for higher eigenstates. This is because the lowest energy states converge for smaller cutoffs.

\begin{figure}[t]\centering
\begin{subfigure}[b]{0.6\textwidth}\includegraphics[width=\textwidth]{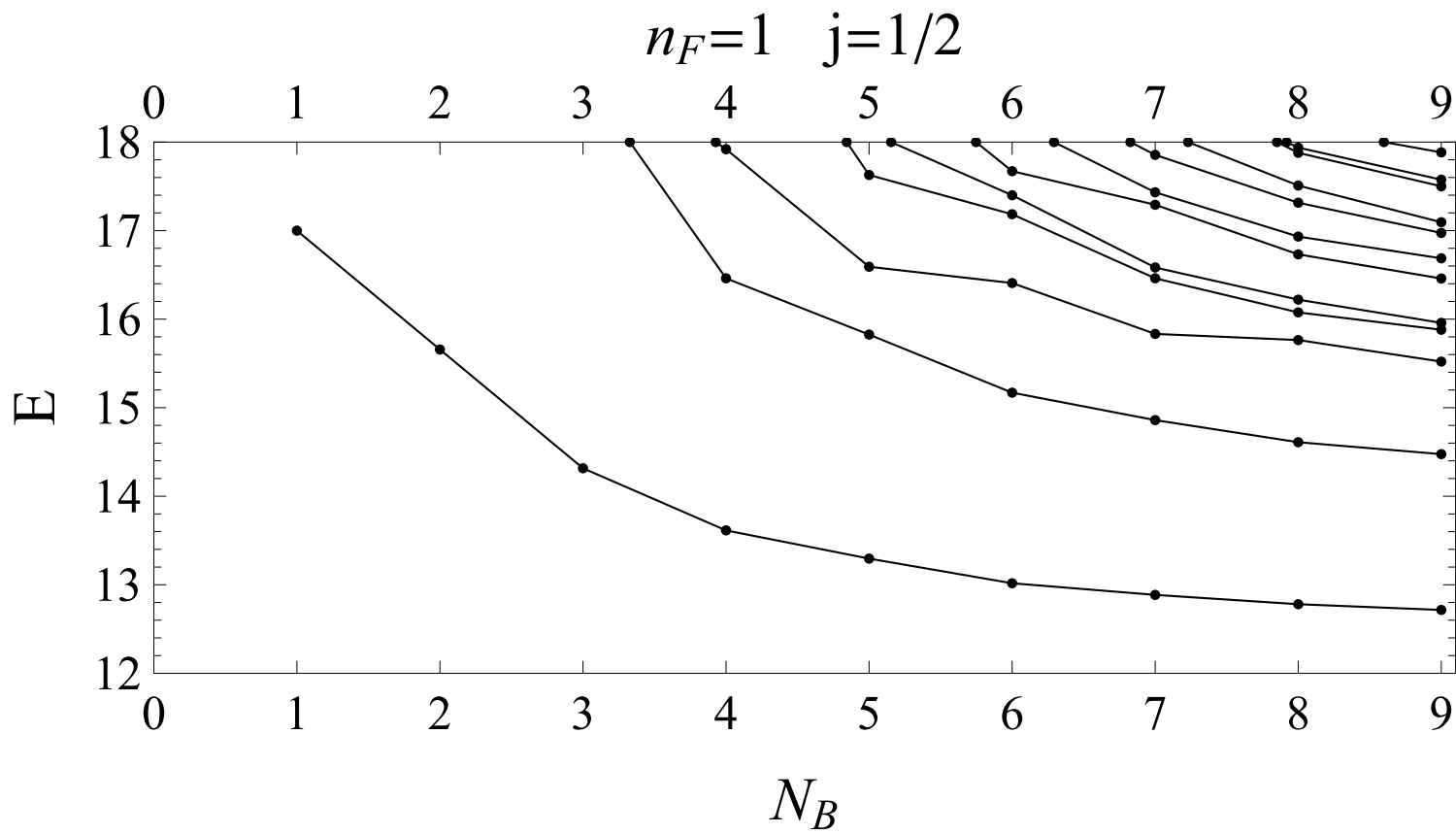}\end{subfigure}
\caption{Spectrum of the $SU(3)$ theory with $n_F=1$, angular momentum $j=\frac{1}{2}$ and $\omega=1$. The third and fourth excited energies for $N_B=9$ are very close. This is the reason why these two states mix and only their orthogonal superpositions give a good approximation to exact energy eigenstates.}
\label{fig:su3_spectrum_nf1}
\end{figure}

The behavior of fractions $q_1(\frac{1}{2},E_i|0,E_1)$, $i=3,4$ in Fig. \ref{fig:fractions_convergence} is more complex. We address this case on a more general ground. Suppose that with the available cutoff a good approximation of a bound state $\ket{n_F,j,m,E}$ is obtained. Denote its (exact) superpartner in the channel $(n_F+1,j'=j\pm\frac{1}{2})$ by $\ket{n_F+1,j',m',E}$. In general, for larger number of fermions the cutoff is smaller and thus the approximation is worse. Therefore, a good approximation of the state $\ket{n_F+1,j',m',E}$ is given rather by some combination of energy eigenstates for a given cutoff
\begin{align}\label{eq:superpartner_decomposition}
\ket{n_F+1,j',m',E}\approx\sum_i c_i\ket{n_F+1,j',m',E_i}_{N_B}.
\end{align}
The supersymmetry fractions $q_{n_F}(j'E_i|jE)$ for cutoff $N_B$ are fully determined by coefficients $c_i$. If $\ket{n_F,j,m,E}$ is vacuum of a supermultiplet then the supersymmetry fraction is
\begin{align}\label{eq:split_fractions}\begin{split}
&q_{n_F}(j',E_i|j,E)=\left(j'+\frac{1}{2}\right)\left|c_i\right|^2.
\end{split}\end{align}
If $\ket{n_F,j,m,E}$ is $\ket{\Omega_+,m}$ or $\ket{\Omega_-,m}$, then
\begin{align}
q_{n_F}(j',E_i|j,E)&=\left(j+\frac{1}{2}\right)\left|c_i\right|^2.
\end{align}
In both cases $q_{n_F}(j',E_i|j,E)$ is the exact supersymmetry fraction multiplied by $|c_i|^2$. Concluding, the supersymmetry fractions $q_{n_F}(j',E_i|j,E)$ add up to the exact value. Therefore, if one finds several fractions $q_{n_F}(j',E_i|j,E)$ which sum up almost to $j+\frac{1}{2}$ or $j'+\frac{1}{2}$ then the superpartner of $\ket{n_F,j,m,E}$ is given by (\ref{eq:superpartner_decomposition}) where $c_i$ are determined up to phases.

Consider again the supersymmetry fractions $q_1(\frac{1}{2},E_i|0,E_1)$ (Fig. \ref{fig:fractions_convergence}). For $N_B$ up to $7$ the fraction $q_1(\frac{1}{2},E_4|0,E_1)$ grows while $q_1(\frac{1}{2},E_3|0,E_1)$ is close to $0$. For $N_B=8$ the two fractions interchange. This is because the energy of the state $\ket{1,\frac{1}{2},E_4}$ decreases faster than the energy of $\ket{1,\frac{1}{2},E_3}$ and labels of energies are swapped at $N_B=8$. For $N_B=9$ the fractions are $0.46$ and $0.47$ and sum up to $0.93$. Based on earlier analysis we conclude that a good approximation to the superpartner $\ket{\Omega_+,m}$ of the state $\ket{0,0,E_1}$ is given by a combination
\begin{align}
\ket{\Omega_+,m}&\approx\sqrt{0.46} e^{i\phi_1}\ket{1,\frac{1}{2},m,E_3}_{N_B=9}+\sqrt{0.47} e^{i\phi_2}\ket{1,\frac{1}{2},m,E_4}_{N_B=9}.
\end{align}
A reason why the two energy states mix for $N_B=9$ is the fact that their energies are close for that cutoff (cf. Fig. \ref{fig:su3_spectrum_nf1}).

Such situation as described above, with several supersymmetry fractions which add up almost to an exact value is in fact very common.
This is because the number of degrees of freedom is large and thus eigenenergies are densely distributed. Therefore, states that correspond to different but close energies mix easily for a finite cutoff. Obviously, these states disentangle for a cutoff which is high enough. Nevertheless, if energies of different supermultiplets are almost degenerate, the needed cutoff is very large.

A more complete set of supersymmetry fractions $q_1(\frac{1}{2},E_i|0,E_k)$ is given in Tab. \ref{tab:susyf00up}. All states from the $(0,0)$ sector are vacua of a $(0,0,E)$ supermultiplet. Therefore, for all of them there should be a supersymmetry fraction equal to $1$ with some states with $(n_F,j)=(1,\frac{1}{2})$. The states from channel $(1,\frac{1}{2})$ belong to supermultiplets which vacua have quantum numbers $(0,0,E),\ (0,1,E)$ or $(1,\frac{1}{2},E)$. That is why some rows in Tab. \ref{tab:susyf00up} are empty.

The first two columns of Tab. \ref{tab:susyf00up} were already analyzed. The third column corresponds to the singlet $\ket{0,0,17.47}$. There are five fractions which correspond to doublets in single fermion sector and add up to $0.83$. Energies of these doublets range from $17.88$ to $18.85$. One can see in Fig. \ref{fig:su3_spectrum_nf1} that for $E\sim18$ energies decrease by $\Delta E\sim1$ when $N_B$ changes from $8$ to $9$. That is, the energies are determined with precision of order 1. It is then natural that the (exact) energy states mix for finite $N_B$ as long as their energies agree within the error.

Analysis of all other sectors is similar. Summary of identified supermultiplets is given in Tab. \ref{tab:spectroscopy}. Values of supersymmetry fractions are also given to indicate how strong is the identification. Clearly, large difference of energies within a supermultiplet means that states did not yet converge and one should not expect large values of fractions. Supersymmetry fractions given in Tab. \ref{tab:spectroscopy} are normalized, i.e. divided by exact values.

All above considerations address only the case of discrete spectrum. This is because the continuous spectrum appears in channels with many fermions. There the density of energy states is higher while the available cutoff is small. This makes the analysis of supersymmetry fractions yet more demanding.

\subsection{Summary}
\begin{figure}[t]\centering
\includegraphics[width=.5\textwidth]{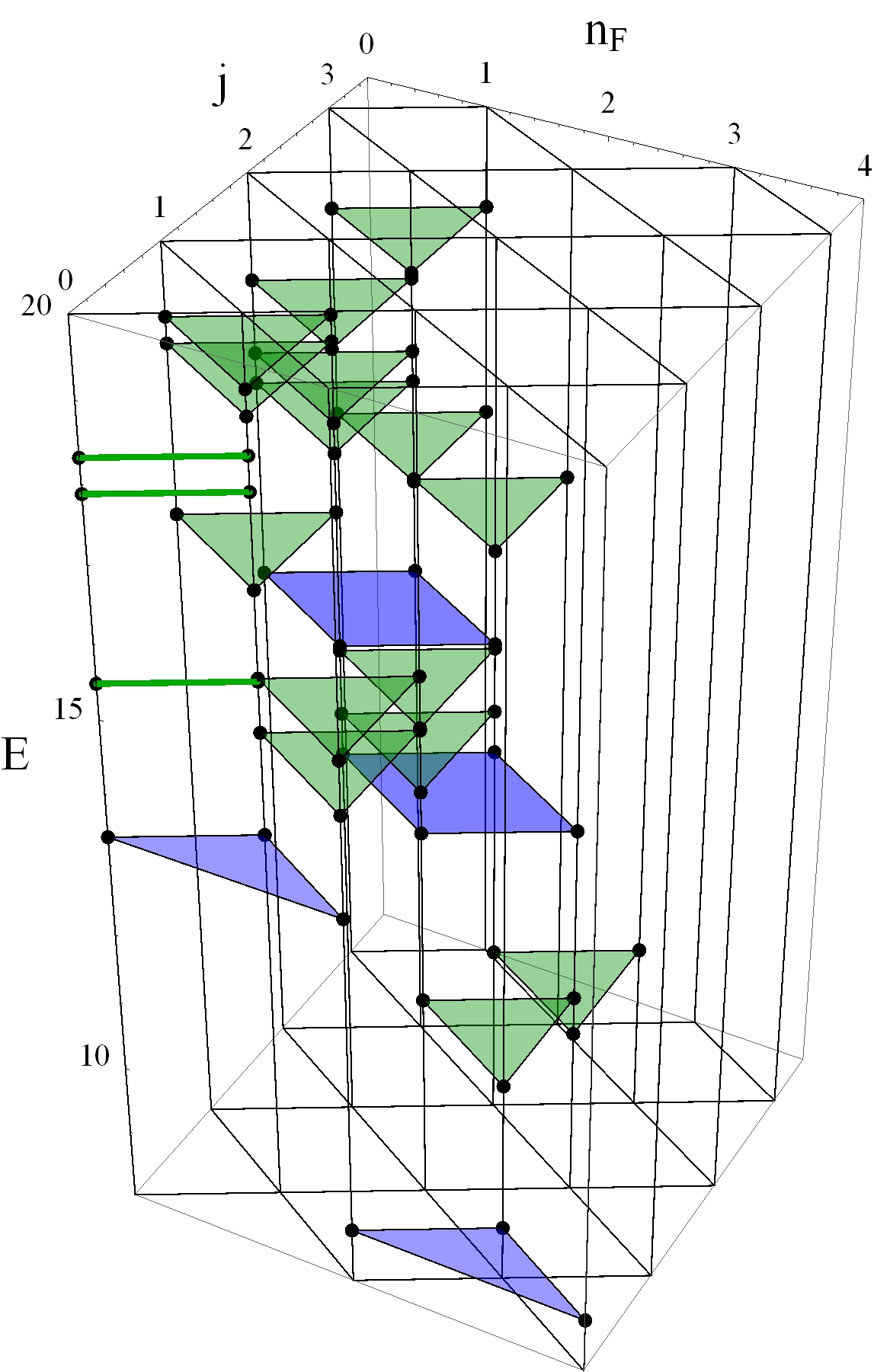}
\caption{All identified supermultiplets. The diamond and two triangles with two vertices at $j=0$ are completely identified multiplets. The other triangles correspond to diamonds with one state unidentified. Lines correspond to triangle multiplets with one vertex not identified.}
\label{fig:overall}
\end{figure}

In Fig. \ref{fig:overall} all identified supermultiplets are shown. Triangles with vertices in sectors $(n_F,0),\ (n_F+1,\frac{1}{2}),\ (n_F+2,0)$ and all diamonds are fully identified supermultiplets. They are marked with blue color. The triangles which have vertices at $(n_F,j)$, $(n_F+1,j-\frac{1}{2})$ and $(n_F+1,j+\frac{1}{2})$ are marked with green color and represent diamonds with one element not identified. Green lines correspond to two states out of three in a multiplet with Clifford vacuum in $(n_F,0)$. The method which was used for finding the energies is variational, so all energies are approximated from above. For this reason all states in Fig. \ref{fig:overall} are shifted to the lowest energy in the corresponding multiplet.

We succeeded to fully identify four supermultiplets (cf. Tab. \ref{tab:spectroscopy}). In general, it is easier to identify states which 'open' the multiplet, i.e. the Clifford vacuum in the sector with $n_F$ fermions and two states with $n_F+1$ fermions. The remaining state which 'closes' the supermultiplet is significantly harder to find. Even if the Clifford vacuum is a low excited state, then its superpartners, which have more fermions and the same energy, are highly excited states (cf. Figs. \ref{fig:su3_spectrum_nf0} - \ref{fig:su3_spectrum_nf_large}). Therefore, one needs higher cutoff before convergence of these states is obtained. However, cutoffs in sectors with more fermions are lower. Therefore, the 'closing' state is rarely well approximated.

\begin{table}\centering
\begin{tabular}{cc|ccccc}
&& \multicolumn{5}{c}{$n_F=0,\ j=0$}    \\
\multirow{16}{*}{\rotatebox[origin=c]{90}{$n_F=1,\ j={1/2}$}}&
$E$&12.6&15.48&17.47&17.88&18.32
\\\hline&12.72&0.99&-&-&-&-
\\&14.48&-&-&-&-&-
\\&15.52&-&0.01&-&-&-
\\&15.88&-&0.46&-&-&-
\\&15.96&-&0.47&0.01&-&-
\\&\vdots
\\&17.88&-&-&0.11&0.03&-
\\&18.26&-&-&-&-&-
\\&18.32&-&-&-&-&-
\\&18.52&-&-&0.2&0.14&0.04
\\&18.62&-&-&0.42&0.03&0.05
\\&18.73&-&-&0.06&0.02&0.01
\\&18.85&-&-&0.05&0.61&0.01
\\&\vdots
\\&19.66&-&-&0.01&0.03&0.11
\\&19.73&-&-&-&-&0.52
\end{tabular}
\caption{Supersymmetry fractions between sectors $(n_F,j)=(0,0)$ and $(1,1/2)$. Only those values which are greater or equal to $0.01$ are given.}
\label{tab:susyf00up}
\end{table}

\begin{table}\centering
\begin{tabular}{c|cccc|cccc}
&\multicolumn{4}{c|}{energy}&\multicolumn{4}{c}{fractions}\\
\hline
$(n_F,j)$& \rotatebox[origin=c]{70}{$(n_F,j)$}& \rotatebox[origin=c]{70}{$(n_F+1,j-\frac{1}{2})$}& \rotatebox[origin=c]{70}{$(n_F+1,j+\frac{1}{2})$}& \rotatebox[origin=c]{70}{$(n_F+2,j)$}& \rotatebox[origin=c]{70}{$\tilde q_{n_F}(j-\frac{1}{2}|j)$}& \rotatebox[origin=c]{70}{$\tilde q_{n_F}(j+\frac{1}{2}|j)$}& \rotatebox[origin=c]{70}{$\tilde q_{n_F+1}(j|j-\frac{1}{2})$}& \rotatebox[origin=c]{70}{$\tilde q_{n_F+1}(j|j+\frac{1}{2})$}\\
\hline
(0,0)&12.6&-&12.72&13.4*&-&.99&-&.92*\\
(0,0)&15.48&-&15.92*&?&-&.93*&-&?\\
(0,0)&17.88&-&18.85&?&-&.61&-&?\\
(0,0)&18.32&-&19.7*&?&-&.63*&-&?\\
(0,1)&16.66&17.5&17.55&?&.77&.76&?&?\\
(0,1)&18.79&20.75&20.78*&?&.53&.52&?&?\\
(0,1)&19.11&20.38*&20.41&?&.54*&.53*&?&?\\
(0,2)&14.89&15.27&15.29&17.06&.9&.95&.75*&.67*\\
(0,2)&17.41&18.47&18.49&?&.72&.79&?&?\\
(0,2)&17.79&18.94&18.98*&?&.66&.71*&?&?\\
(0,2)&18.7&20.8&20.83*&?&.6&.63*&?&?\\
(0,3)&16.1&16.75&16.8&?&.86&.9&?&?\\
(0,3)&18.77&20.9*&21.02&?&.5*&.66&?&?\\
(1,1/2)&14.48&15.67&15.62&?&.6&.65&?&?\\
(1,1/2)&15.52&16.93*&16.72&?&.52&.75&?&?\\
(1,3/2)&13.39&13.86&13.97&14.96&.83*&.88*&.2*&.18\\
(1,3/2)&13.96&14.97&15.05&?&.73*&.8*&?&?\\
(1,3/2)&14.83&15.81&16.14&?&.71*&.71*&?&?\\
(1,5/2)&16.13&17.61&17.94&?&.49&.57*&?&?\\
(2,0)&8.8&-&9.84&12.74&-&.83&-&.58\\
(2,1)&10.97&13.&12.93&?&.62&.71*&?&?\\
(2,2)&10.43&12.68&12.72&?&.59*&.59&?&?
\end{tabular}
\caption{Energies and supersymmetry fractions in identified supermultiplets. Fractions are divided by the value they should take in continuum. The closer the normalized supersymmetry fraction is to one, the higher is the quality of identification of a supermultiplet. Stars mean that the fraction is summed over two states and energies are averaged. Question marks stand for unidentified states and corresponding supersymmetry fractions in a supermultiplet. Dashes stand for states which do not appear in a supermultiplet.}
\label{tab:spectroscopy}
\end{table}

\section{Summary}
In this paper the supersymmetric Yang-Mills quantum mechanics in $D=4$ with the $SU(3)$ gauge group was studied. The cut Fock space method was used to construct matrix representations of relevant operators of the theory. They served to find energies of the model and to analyze supersymmetric structure of SYMQM.

The physical Hilbert space consists of $SU(3)$ singlets, which are constructed by acting with so called bricks on the empty state. We found that there are 786 independent bricks. Next, we presented a recursive algorithm to construct matrix elements of an arbitrary operator. The number of fermions is limited by $16$ and it is necessary to take only $n_F\leq8$ because the Hamiltonian is symmetric under the particle-hole transformation. In each fermionic sector we set a cutoff $N_B$ on the total number of bosonic excitations.  A program based on the presented algorithm was used to construct matrices of the Hamiltonian, total angular momentum and supercharges. Next, eigenenergies with definite angular momentum were computed.

In Section \ref{ch:spectrum} the spectrum was analyzed. Our results in the purely bosonic part of the Hilbert space were compared with \cite{Weisz} where no fermions were considered. Very good agreement was found (cf. Tab. \ref{tab:Weisz_comparison}). Results in sectors with fermions are original. Dependence of the eigenenergies on the cutoff $N_B$ is crucial for identifying the type of the spectrum. Discrete energy levels converge very fast to the exact value. On the contrary, it was argued in \cite{Campostrini,Kotanski} that eigenenergies corresponding to the continuous spectrum should satisfy the scaling relation (\ref{eq:continuous_scaling}) $E_k\propto 1/N_B$, where $k$ is the label of the energy. Therefore, the scaled energy $N_B\times E_k$ should approach a constant when $N_B\to\infty$.

For $n_F\leq4$ the scaled energies grow approximately linearly with $N_B$ (cf. Fig. \ref{fig:su3_scaling}). Therefore, the spectrum is discrete. Although the potential has flat valleys, the transverse oscillations give rise to an effective potential. It prevents wavefunctions from propagating in these directions.
For $n_F=6,7,8$ scaled energies seem to be approximately constant, which suggests that the spectrum is continuous. That means that supersymmetric cancelations of bosonic and fermionic degrees of freedom occur and flat valleys of the potential are open. Still, it is possible, and in fact required by supersymmetry, that there are discrete energies immersed in the continuous spectrum. Because of the particle-hole symmetry, the energy levels and nature of the spectrum for $8<n_F\leq16$ is the same as for $16-n_F$ fermions. The sector $n_F=5$ (and thus also $n_F=11$) is disputable. For a more detailed picture concerning the type of spectrum a higher cutoff would be needed.

In Section \ref{ch:supersymmetric_multiplets} we studied supersymmetric structure of the model. In the continuum limit all states are grouped in supermultiplets. However, finite cutoff breaks the supersymmetry. We used supersymmetry fractions $q_{n_F}\left(j'E'|jE\right)$ to identify supermultiplets and to check how much the supersymmetry is broken. Values of supersymmetry fractions are known when the supersymmetry is exact: $q_{n_F}\left(j'E'|jE\right)$ equals $j$ or $j+\frac{1}{2}$ if the two corresponding states are in the same multiplet and zero otherwise. We found that at least for some low energy states the supersymmetry fractions converge to their exact values (cf. Fig. \ref{fig:fractions_convergence}). They were used to identify supermultiplets. Four complete supermultiplets were fully identified. More supermultiplets were found with one state in the channel with highest $n_F$ missing. Complete list of identified supermultiplets is given in Tab. \ref{tab:spectroscopy} and an overview is presented in Fig. \ref{fig:overall}. We conclude that within the available cutoff, some supersymmetry is already restored at low energies and small number of fermions.

\section{Acknowledgements}
This article partly summarizes PhD thesis 'Four dimensional supersymmetric Yang-Mills quantum mechanics with three colors' written by the author at the Jagiellonian University under the supervision of prof. J. Wosiek.

This work was partially supported through NCN grant nr 2011/03/D/ST2/01932.

This work was supported by Foundation for Polish Science MPD Programme co--financed by the European Regional Development Fund, agreement no. MPD/2009/6.

The research was carried out with the supercomputer ,,Deszno'' purchased thanks to the financial support of the European Regional Development Fund in the framework of the Polish Innovation Economy Operational Program (contract no. POIG. 02.01.00-12-023/08).


\begin{thebibliography}{10}

\bibitem{Claudson}
M.~Claudson and M.~B. Halpern.
\newblock {Supersymmetric ground state wave functions}.
\newblock {\em Nucl.Phys.}, B250:689, 1985.

\bibitem{Ferrara}
S.~Ferrara and B.~Zumino.
\newblock {Supergauge invariant Yang-Mills theories}.
\newblock {\em Nucl.Phys.}, B79:413, 1974.

\bibitem{Brink}
L.~Brink, J.~H. Schwarz, and J.~Scherk.
\newblock {Supersymmetric Yang-Mills theories}.
\newblock {\em Nucl.Phys.}, B121:77, 1977.

\bibitem{Bjorken}
J.D. Bjorken.
\newblock {Elements of Quantum Chronodynamics}.
\newblock 1979.

\bibitem{Munster}
M.~Luscher and G.~Munster.
\newblock {Weak coupling expansion of the low lying energy values in the SU(2)
  gauge theory on a torus}.
\newblock {\em Nucl.Phys.}, B232:445, 1984.

\bibitem{Weisz}
P.~Weisz and V.~Ziemann.
\newblock {Weak coupling expansions of the low lying energy values in SU(3)
  gauge theory on a torus}.
\newblock {\em Nucl.Phys.}, B284:157, 1987.

\bibitem{Ziemann_phd}
V.~Ziemann.
\newblock {\em {Qualitative Untersuchung des niedrig liegenden Spektrums reiner
  Yang-Mills Theorien im endlichen Volumen mit besonderer Ber\"ucksichtigung
  von SU(3)}}.
\newblock PhD thesis, {Universit\"at Hamburg}, June 1986.

\bibitem{Koller1}
J.~Koller and P.~van Baal.
\newblock {A rigorous nonperturbative result for the glueball mass and electric
  flux energy in a finite volume}.
\newblock {\em Nucl.Phys.}, B273:387, 1986.

\bibitem{Koller2}
J.~Koller and P.~van Baal.
\newblock {A nonperturbative analysis in finite volume gauge theory}.
\newblock {\em Nucl.Phys.}, B302:1, 1988.

\bibitem{Hoppe}
J.~Hoppe.
\newblock {Quantum theory of a massless relativistic surface and a
  two-dimensional bound state problem}.
\newblock {\em unpublished}, http://hdl.handle.net/1721.1/15717.

\bibitem{Bergshoeff}
E.~Bergshoeff, E.~Sezgin, and P.K. Townsend.
\newblock {Supermembranes and eleven-dimensional supergravity}.
\newblock {\em Phys.Lett.}, B189:75--78, 1987.

\bibitem{deWit}
B.~de~Wit, J.~Hoppe, and H.~Nicolai.
\newblock {On the quantum mechanics of supermembranes}.
\newblock {\em Nucl.Phys.}, B305:545, 1988.

\bibitem{Nicolai}
B.~de~Wit, M.~Luscher, and H.~Nicolai.
\newblock {The supermembrane is unstable}.
\newblock {\em Nucl.Phys.}, B320:135, 1989.

\bibitem{Helling}
H.~Nicolai and R.~Helling.
\newblock {Supermembranes and M(atrix) theory}.
\newblock 1998, hep-th/9809103.

\bibitem{BFSS}
T.~Banks, W.~Fischler, S.H. Shenker, and L.~Susskind.
\newblock {M theory as a matrix model: A Conjecture}.
\newblock {\em Phys.Rev.}, D55:5112--5128, 1997, hep-th/9610043.

\bibitem{Becker}
Katrin Becker and Melanie Becker.
\newblock {A two loop test of M(atrix) theory}.
\newblock {\em Nucl.Phys.}, B506:48--60, 1997, hep-th/9705091.

\bibitem{Porrati}
M.~Porrati and A.~Rozenberg.
\newblock {Bound states at threshold in supersymmetric quantum mechanics}.
\newblock {\em Nucl.Phys.}, B515:184--202, 1998, hep-th/9708119.

\bibitem{Taylor}
W.~Taylor.
\newblock {M(atrix) theory: Matrix quantum mechanics as a fundamental theory}.
\newblock {\em Rev.Mod.Phys.}, 73:419--462, 2001, hep-th/0101126.

\bibitem{Danielsson}
U.~H. Danielsson, G.~Ferretti, and B.~Sundborg.
\newblock {D particle dynamics and bound states}.
\newblock {\em Int.J.Mod.Phys.}, A11:5463--5478, 1996, hep-th/9603081.

\bibitem{Halpern}
M.B. Halpern and C.~Schwartz.
\newblock {Asymptotic search for ground states of SU(2) matrix theory}.
\newblock {\em Int.J.Mod.Phys.}, A13:4367--4408, 1998, hep-th/9712133.

\bibitem{Catterall}
S.~Catterall and T.~Wiseman.
\newblock {Black hole thermodynamics from simulations of lattice Yang-Mills
  theory}.
\newblock {\em Phys.Rev.}, D78:041502, 2008, arXiv:0803.4273[hep-th].

\bibitem{Wiseman}
S.~Catterall and T.~Wiseman.
\newblock {Extracting black hole physics from the lattice}.
\newblock {\em JHEP}, 1004:077, 2010, arXiv:0909.4947 [hep-th].

\bibitem{Anagnostopoulos}
K.~N. Anagnostopoulos, M.~Hanada, J.~Nishimura, and S.~Takeuchi.
\newblock {Monte Carlo studies of supersymmetric matrix quantum mechanics with
  sixteen supercharges at finite temperature}.
\newblock {\em Phys.Rev.Lett.}, 100:021601, 2008, arXiv:0707.4454 [hep-th].

\bibitem{Hanada}
M.~Hanada, Y.~Hyakutake, G.~Ishiki, and J.~Nishimura.
\newblock {Holographic description of quantum black hole on a computer}.
\newblock 2013, arXiv:1311.5607 [hep-th].

\bibitem{Nishimura}
M.~Hanada, J.~Nishimura, and S.~Takeuchi.
\newblock {Non-lattice simulation for supersymmetric gauge theories in one
  dimension}.
\newblock {\em Phys.Rev.Lett.}, 99:161602, 2007, arXiv:0706.1647 [hep-lat].

\bibitem{Janik}
R.A. Janik and J.~Wosiek.
\newblock {Towards the matrix model of M theory on a lattice}.
\newblock {\em Acta Phys.Polon.}, B32:2143--2154, 2001, hep-th/0003121.

\bibitem{Wosiek}
J.~Wosiek.
\newblock {Spectra of supersymmetric Yang-Mills quantum mechanics}.
\newblock {\em Nucl.Phys.}, B644:85--112, 2002, hep-th/0203116.

\bibitem{Campostrini}
M.~Campostrini and J.~Wosiek.
\newblock {High precision study of the structure of D=4 supersymmetric
  Yang-Mills quantum mechanics}.
\newblock {\em Nucl.Phys.}, B703:454--498, 2004, hep-th/0407021.

\bibitem{Trzetrzelewski_susyd2}
M.~Trzetrzelewski.
\newblock {Large N behavior of two dimensional supersymmetric Yang-Mills
  quantum mechanics}.
\newblock {\em J.Math.Phys.}, 48:012302, 2007, hep-th/0608147.

\bibitem{Korcyl}
P.~Korcyl.
\newblock {Exact solutions to D=2 Supersymmetric Yang-Mills Quantum Mechanics
  with SU(3) gauge group}.
\newblock {\em Acta Phys.Polon.Supp.}, 2:623, 2009, arXiv:0911.2152 [hep-th].

\bibitem{KorcylN}
P.~Korcyl.
\newblock {Solutions of D=2 supersymmetric Yang-Mills quantum mechanics with
  SU(N) gauge group}.
\newblock {\em J.Math.Phys.}, 52:052105, 2011, arXiv: 1101.0591 [math-ph].

\bibitem{vanBaal}
P.~van Baal.
\newblock {The Witten index beyond the adiabatic approximation}.
\newblock 2001, hep-th/0112072.

\bibitem{Kotanski}
J.~Kotanski.
\newblock {Energy spectrum and wave-functions of four-dimensional
  Supersymmetric Yang-Mills Quantum Mechanics for very high cut-offs}.
\newblock {\em Acta Phys.Polon.}, B37:2813--2838, 2006, hep-th/0607012.

\bibitem{Kotanski2}
Jan Kotanski.
\newblock {Virial theorem for four-dimensional supersymmetric Yang-Mills
  quantum mechanics with SU(2) gauge group}.
\newblock {\em Acta Phys.Polon.}, B37:3659--3666, 2006, hep-th/0610091.

\bibitem{Itzykson}
C.~Itzykson and J.B. Zuber.
\newblock {\em {Quantum Field Theory}}.
\newblock {McGraw-Hill}, 1980.

\bibitem{Weinberg}
S.~Weinberg.
\newblock {\em {The quantum theory of fields. Vol. 3: Supersymmetry}}.
\newblock Cambridge University Press, 2000.

\bibitem{Dancoff}
S.M. Dancoff.
\newblock {Nonadiabatic meson theory of nuclear forces}.
\newblock {\em Phys.Rev.}, 78:382--385, 1950.

\bibitem{Brodsky}
Hans~Christian Pauli and Stanley~J. Brodsky.
\newblock {Discretized Light Cone Quantization: Solution to a Field Theory in
  One Space One Time Dimensions}.
\newblock {\em Phys.Rev.}, D32:2001, 1985.

\bibitem{Trzetrzelewski_spectra}
M.~Trzetrzelewski and J.~Wosiek.
\newblock {Quantum systems in a cut Fock space}.
\newblock {\em Acta Phys.Polon.}, B35:1615--1624, 2004, hep-th/0308007.

\bibitem{Ambrozinski}
Z.~Ambrozinski and J.~Wosiek.
\newblock {Resumming not summable perturbative series}.
\newblock {\em Acta Phys.Polon.}, B44(1):49--58, 2013.

\bibitem{Ambrozinski_cosine}
Z.~Ambrozinski.
\newblock {Tunneling in cosine potential with periodic boundary conditions}.
\newblock {\em Acta Phys.Polon.}, B44:1261--1272, 2013, arXiv: 1303.0708
  [quant-ph].

\bibitem{Trzetrzelewski_trees}
M.~Trzetrzelewski.
\newblock {Reduction of su(N) loop tensors to trees}.
\newblock {\em J.Math.Phys.}, 46:103512, 2005, math-ph/0505084.

\bibitem{Procesi}
M.~Bresan, C.~Procesi, and S.~Spenko.
\newblock {Quasi-identities on matrices and the Cayley-Hamilton polynomial}.
\newblock arXiv: 1212.4597.

\bibitem{Macfarlane}
A.J. Macfarlane, A.~Sudbery, and P.~Weisz.
\newblock {Explicit representations of chiral invariant lagrangian theories of
  hadron dynamics}.
\newblock {\em Proc.Roy.Soc.Lond.}, 314:217--250, 1970.

\bibitem{doktorat}
Z.~Ambrozi\'nski.
\newblock {\em {Four dimensional supersymmetric Yang-Mills quantum mechanics
  with three colors}}.
\newblock PhD thesis, {Jagiellonian University}, June 2014, arXiv:1408.2655
  [hep-th].

\bibitem{Trzetrzelewski_number}
M.~Trzetrzelewski.
\newblock {The Number of gauge singlets in supersymmetric Yang-Mills quantum
  mechanics}.
\newblock {\em Phys.Rev.}, D76:085012, 2007, arXiv:0708.2946 [hep-th].

\bibitem{Trzetrzelewski1}
M.~Trzetrzelewski.
\newblock {Quantum mechanics in a cut Fock space}.
\newblock {\em Acta Phys.Polon.}, B35:2393--2416, 2004, hep-th/0407059.

\bibitem{Simon}
B.~Simon.
\newblock {Some quantum operators with discrete spectrum but clasically
  continuous spectrum}.
\newblock {\em Annals Phys.}, 146:209--220, 1983.

\bibitem{Korcyl_effective}
P.~Korcyl.
\newblock {Classical trajectories and quantum supersymmetry}.
\newblock {\em Phys.Rev.}, D74:115012, 2006, hep-th/0610105.

\bibitem{Wess}
J.~Wess and J.~Bagger.
\newblock {\em {Supersymmetry and supergravity}}.
\newblock Princeton University Press, 1983.

\end{thebibliography}

\end{document}